\documentclass[aps, prx,twocolumn,longbibliography,superscriptaddress,showpacs,amsmath,amssymb,floatfix]{revtex4}
\usepackage[latin9]{inputenc}
\usepackage{amssymb}
\usepackage{graphicx}
\usepackage{amsmath}
\usepackage{color}
\usepackage{mathrsfs}
\usepackage{float}
\usepackage{indentfirst}
\usepackage{mathrsfs}
\usepackage{float}
\usepackage{indentfirst}
\usepackage{textcomp}
\usepackage{comment}
\usepackage{mathtools}
\usepackage{natbib,hyperref}
\usepackage{soul}

\begin{document}

\title{Combination of Tensor Network States and Green's function Monte Carlo}

\author{Mingpu Qin}
\affiliation{Key Laboratory of Artificial Structures and Quantum Control, School of Physics and Astronomy, Shanghai Jiao Tong University, Shanghai 200240, China}

\begin{abstract}
We propose an approach to study the ground state of quantum many-body systems in which Tensor Network States (TNS), specifically  Projected Entangled Pair States (PEPS), and Green's function Monte Carlo (GFMC) are combined. PEPS, by design, encode the area law which governs the scaling of entanglement entropy in quantum systems with short range interactions but are hindered by the high computational complexity scaling with bond dimension (D). GFMC is a highly efficient method, but it usually suffers from the infamous negative sign problem which can be avoided by the fixed node approximation in which a guiding wave function is utilized to modify the sampling process. The trade-off for the absence of negative sign problem is the introduction of systematic error by guiding wave function. In this work, we combine these two methods, PEPS and GFMC, to take advantage of both of them. PEPS are very accurate variational wave functions, while at the same time, only contractions of single-layer tensor network are necessary in GFMC, which reduces the cost substantially. Moreover, energy obtained in GFMC is guaranteed to be variational and lower than the variational energy of the guiding PEPS wave function. Benchmark results of $J_1$-$J_2$ Heisenberg model on square lattice are provided.       
\end{abstract}

\maketitle


One of the most challenging tasks in condensed matter physics is to understand the many-body effect in strongly correlated quantum systems, in which
numerous exotic phenomena emerge \cite{Bednorz1986,PhysRevLett.48.1559,PhysRevLett.50.1395,RevModPhys.89.025003}. 
Because exact solution for strongly
correlated system is rare \cite{PhysRevLett.20.1445}, most studies of these systems rely on numerical tools nowadays.
Density Matrix Renormalization Group (DMRG) \cite{PhysRevLett.69.2863,PhysRevB.48.10345} is one of the most successful methods in the study of strongly correlated systems. DMRG
is extremely accurate for one-dimensional (1D) quantum systems \cite{RevModPhys.77.259, SCHOLLWOCK201196}. Shortly after the introduction of
DMRG in 1992 \cite{PhysRevLett.69.2863},
it was realized that the underlying wave functions are Matrix Product States (MPS) \cite{PhysRevLett.75.3537},
which can be viewed as a generalization of the seminal AKLT state \cite{PhysRevLett.59.799}.  
The concept of MPS can be traced back at least to 1968 \cite{doi:10.1063/1.1664623}.
It was found that MPS capture the entanglement structure of
the ground state of 1D
quantum systems and this results in
the high accuracy of DMRG \cite{RevModPhys.82.277}. The adoption of concepts from the field of quantum information, entanglement for example, have
inspired the
development of the Tensor Network States (TNS) \cite{doi:10.1080/14789940801912366}.
These advances provide us useful tools to both identify \cite{ORUS2014117} and classify \cite{PhysRevB.83.035107,PhysRevB.84.165139} quantum phases.  


DMRG can also provide accurate results for systems on narrow cylinders with large enough bond dimension \cite{doi:10.1146/annurev-conmatphys-020911-125018, Yan1173,PhysRevX.5.041041,Zheng1155, PhysRevX.10.031016} but has difficulty for real 2D system \cite{PhysRevB.49.9214}.  
A natural generalization of MPS to
2D, Projected Entangled Pair States (PEPS) \cite{2004cond.mat..7066V}, can
overcome the difficulty \cite{PhysRevLett.101.090603,PhysRevLett.101.250602,PhysRevLett.107.215301,PhysRevLett.113.046402,PhysRevLett.118.137202}.
It can be proved that the entanglement entropy in PEPS satisfies the area law \cite{2004cond.mat..7066V} which is required to faithfully represent the ground state of
2D quantum systems \cite{RevModPhys.82.277} (when a Fermi surface is present \cite{PhysRevLett.96.100503}, there
is a logarithm correction to area law for the ground state, which PEPS fails to capture.). 
However, in contrast to $D^3$ scaling of complexity in MPS, the computational cost is as 
high as $D^{12}$ \cite{PhysRevLett.101.250602,PhysRevB.80.094403,PhysRevB.92.035142} in PEPS. The heavily scaling of computational resource with bond dimension in PEPS hampers the reach to large bond dimension
which is essential to resolve possible competing states in
the low energy manifold of certain strongly correlated systems, e.g., the anti-ferromagnetic Heisenberg model on Kagome lattice  \cite{PhysRevLett.98.117205,Yan1173,PhysRevLett.109.067201,PhysRevB.87.060405,PhysRevLett.118.137202,PhysRevB.95.235107}.

Quantum Monte Carlo (QMC) \cite{PhysRevD.24.2278,PhysRevB.80.094403,Assaad2008,Zhang2004} is a widely used methodology in the study of strongly correlated many-body systems. 
In general, the computational complexity in QMC scales algebraically with system size which makes it an efficient approach.
However, with few exception \cite{PhysRevB.80.094403}, the direct application of Monte Carlo method in many-body systems
suffers from the infamous negative sign problem \cite{PhysRevB.41.9301,PhysRevLett.94.170201}.
One strategy to overcome the negative sign problem is to take advantage of the trade-off between variance and bias, which is the principle behind fixed node approximation in
Green's function Monte Carlo (GFMC) \cite{PhysRev.128.1791,PhysRevLett.45.566} (also named as diffusion Monte Carlo (DMC) \cite{RevModPhys.73.33} in the literature),
and constrained path approximation in auxiliary field quantum Monte Carlo \cite{PhysRevB.55.7464}.
The price
to pay is the introduction of systematic error or bias in the result. 
Empirically, different forms of guiding \cite{PhysRevB.44.9410,PhysRevLett.72.2442,PhysRevB.87.060405,PhysRevLett.82.3899} or trial wave functions \cite{PhysRevB.78.165101,PhysRevB.94.235119} can be chosen to give high accuracy for certain systems.

\begin{figure}[t]
	\includegraphics[width=80mm]{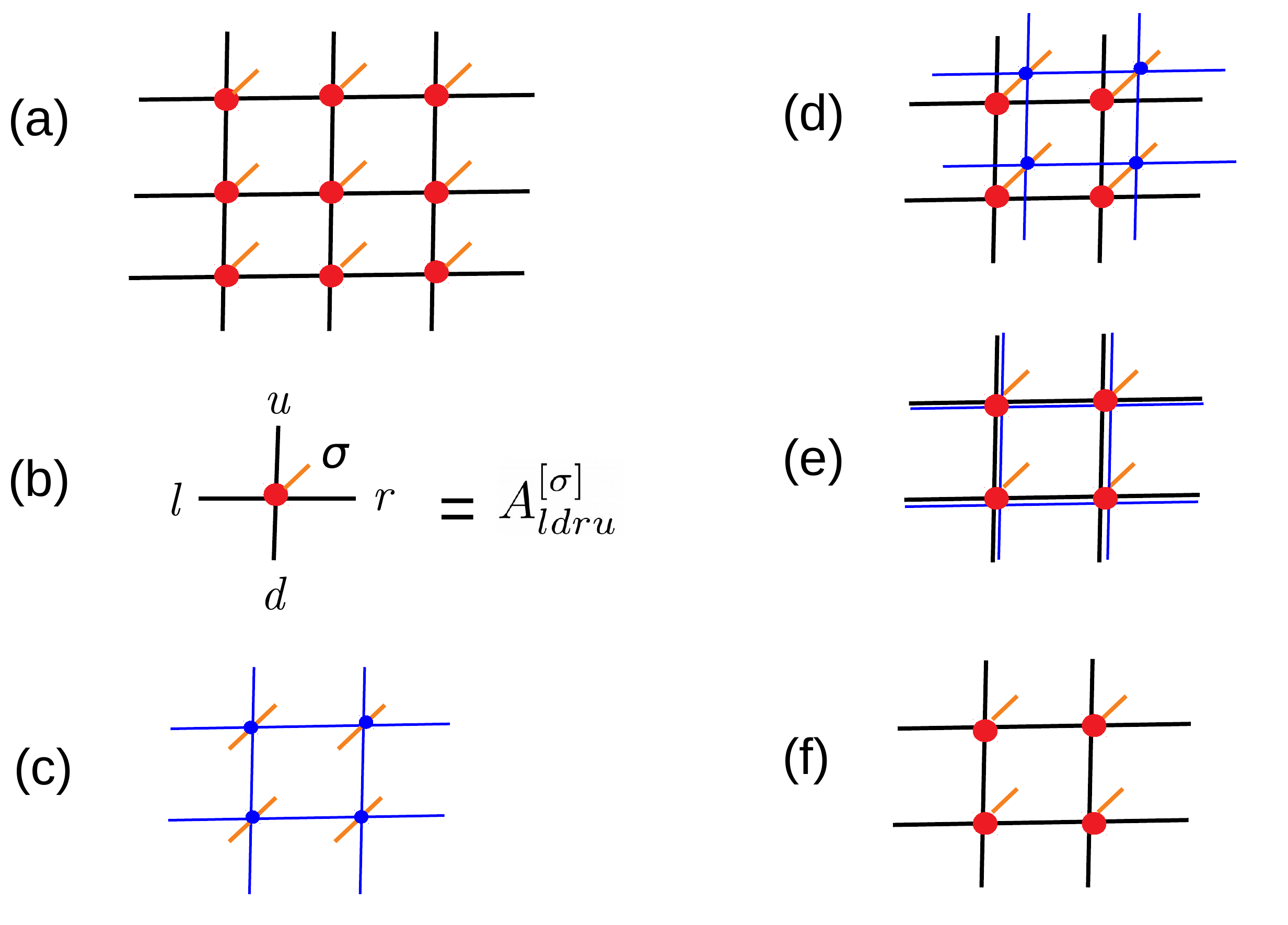}
	\caption{PEPS and the optimization process. (a) shows a PEPS on a square lattice, where there is a local tensor on each vertex. (b) is the local tensor. The wave function is obtained by contracting all the
		auxiliary indexes. (c) shows PEPO for a plaqueate. (d) the application of a PEPO to PEPS in a plaqueate. (e) the bond dimension of PEPS
		is increased after the application of a PEPO, (f) the bond dimension is truncated back to the original value as in (a).
	}
	\label{PEPS}
\end{figure}

In this work, we combine PEPS and GFMC to take advantage of both of them.
We take PEPS as guiding wave function in GFMC calculation.
As we will discuss late, in our method, only contraction of single-layer tensor network is required, which highly reduces the computational complexity in the PEPS part. At the
same time, PEPS are very accurate variational wave functions, with which the systematic error can be reduced in GFMC.

{\em Models --} For concreteness, we take the $S=1/2$ $J_1$-$J_2$ Heisenberg model on square lattice as an example to describe the method. The Hamiltonian is as follows:

\begin{equation}
H=J_{1}\sum_{\langle i,j\rangle}S_{i}S_{j}+J_{2}\sum_{\langle\langle i,j\rangle\rangle}S_{i}S_{j}
\label{Ham}
\end{equation}
where $S_i$ is the spin operator on site $i$. $\langle i,j \rangle$ and $\langle\langle i,j \rangle\rangle$ represent nearest and next nearest neighboring interactions
respectively. We consider anti-ferromagnetic interactions for both $J_1$ and $J_2$, and set $J_1$ as the energy unit. 
For simplicity, we study system on square lattice with open boundary conditions (OBC). The $J_1$-$J_2$ Heisenberg model is
widely investigated in the exploration of the frustration effect in quantum systems \cite{PhysRevB.38.9335,PhysRevLett.63.2148,PhysRevLett.84.3173}. When $J_2$ is absent,
the model can be solved with QMC without suffering  
from the negative sign
problem \cite{PhysRevB.56.11678,PhysRevB.83.155120,PhysRevB.82.024407}. 
The ground state is known to have the long-range anti-ferromagnetic (AF) order, i.e., the Neel order.
In the other
limit when $J_1 = 0$,
the system decouples into two independent square lattices and an infinitesimal $J_1$ can drive the ground state into the striped AF order. Between these two
limits, the nature of the
ground state is still under extensive debates \cite{PhysRevB.74.144422,PhysRevB.79.195119,PhysRevB.85.094407,PhysRevB.86.024424,PhysRevB.86.045115,PhysRevB.88.060402,PhysRevLett.113.027201,doi:10.7566/JPSJ.84.024720,PhysRevLett.121.107202} in the vicinity
of $J_2/J_1 = 0.5$ where the 
system is maximally frustrated in the sense that in the 
classical limit, the $J_1$-$J_2$ Ising model has macroscopic degenerate ground state. 

\begin{figure}[t]
	\includegraphics[width=80mm]{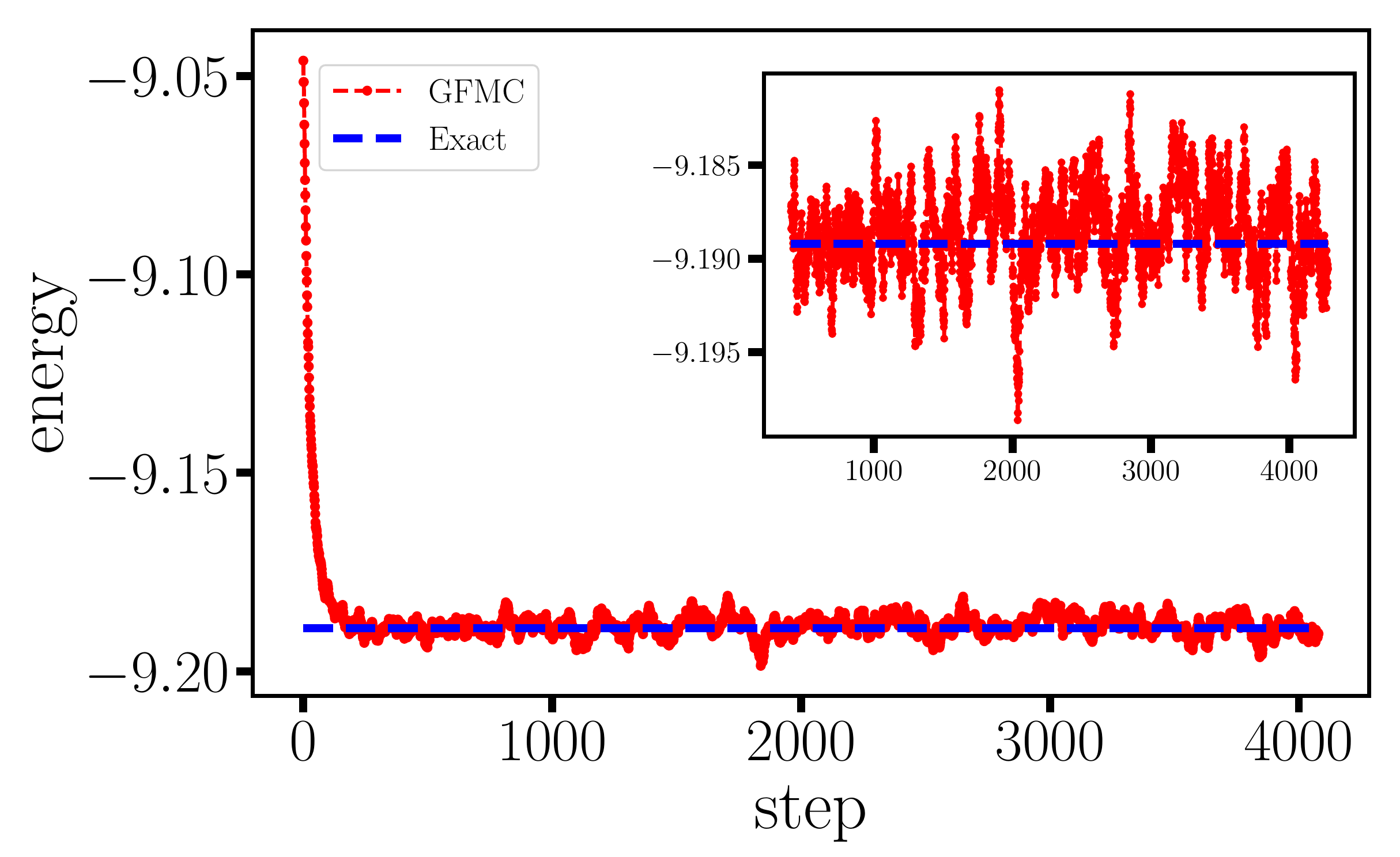}
	\caption{Energy versus step in the GFMC calculation for a $4 \times 4, J_2 = 0$ system with OBC. A PEPS with bond dimension $D = 4$ is used as the guiding function. The energy at step $0$ is the variation energy of the $D=4$ PEPS which is $-9.034333$. There is no sign problem in this case so GFMC should give numerical exact energy. The exact energy is 
		$-9.189207$
		as indicated by the blue dashed line, while the GFMC energy is $-9.189(1)$. The inset shows a zoom of the energy after equilibrium. 
	}
	\label{4_4_j2_0}
\end{figure}
  


{\em Projected Entangled Pair States --} 
To construct a PEPS,
we put a rank five tensor, $A_{ldru}^{[\sigma]}$ on each vertex of the square lattice (see Fig.~\ref{PEPS} (b)), where $\sigma$ is the physical degree of freedom with dimension $d$, while $l,d,r,u$ are the auxiliary degree of freedoms with dimension $D$ \cite{2004cond.mat..7066V}. Contracting the auxiliary degree of freedoms gives the PEPS wave function
\begin{equation}
\psi=\sum_{\{\sigma\}}Tr(A^{[\sigma_{0}]}A^{[\sigma_{1}]}\dots A^{[\sigma_{N-1}]})|\sigma_{0},\sigma_{1},\dots,\sigma_{N-1}\rangle
\end{equation}
where the trace means contraction of tensors.
To obtain the ground state of the $J_1-J_2$ Heisenberg model in Eq(\ref{Ham}), we apply the imaginary time projection operator $\exp(-\tau H)$
repeatedly to an initial PEPS till convergence. Same as in Trotter-Suzuki decomposition \cite{10.1143/PTP.56.1454}, we first divide the lattice into four
groups of plaquettes in a way that the projection operators of each plaquette within a group are independent or commute with each other.  
In each group, a Projected Entangled Pair Operator (PEPO) (see Fig.~\ref{PEPS} (c)) for the projection operator within a single plaquette is constructed and are applied to
the initial PEPS simultaneously. This procedure is carried out recurrently for the four groups (see Fig.~\ref{PEPS} (d)). Because the bond dimension increases after the
 application of PEPO \cite{Pirvu_2010} (see Fig.~\ref{PEPS} (e)), we need
 to truncate it back to the original value $D$ to make the
 calculation under-control (see Fig.~\ref{PEPS} (f)).
 This is done in a variational fashion \cite{PhysRevB.98.085155}, 
The whole procedure is a realization of the cluster update approach \cite{2011arXiv1110.4362W} which is a generalization of simple update \cite{PhysRevLett.101.090603}.
The most time-consuming part of this 
approach is the calculation of physical quantities after the PEPS are optimized, which requires contraction of double-layer tensor networks with bond-dimension $D^2$. As we will discuss late, when
 combining PEPS with GFMC, we only need to calculate single-layer tensor networks with bond-dimension $D$, which reduces the complexity substantially. We notice the
existence of single-layer like algorithm in the literature \cite{PhysRevB.96.045128}. In more sophisticated full update scheme of PEPS, the contraction of double-layer tensor network
is also needed in the optimization process \cite{PhysRevLett.101.250602,PhysRevB.81.165104}.
    
    \begin{figure}[t]
    	\includegraphics[width=80mm]{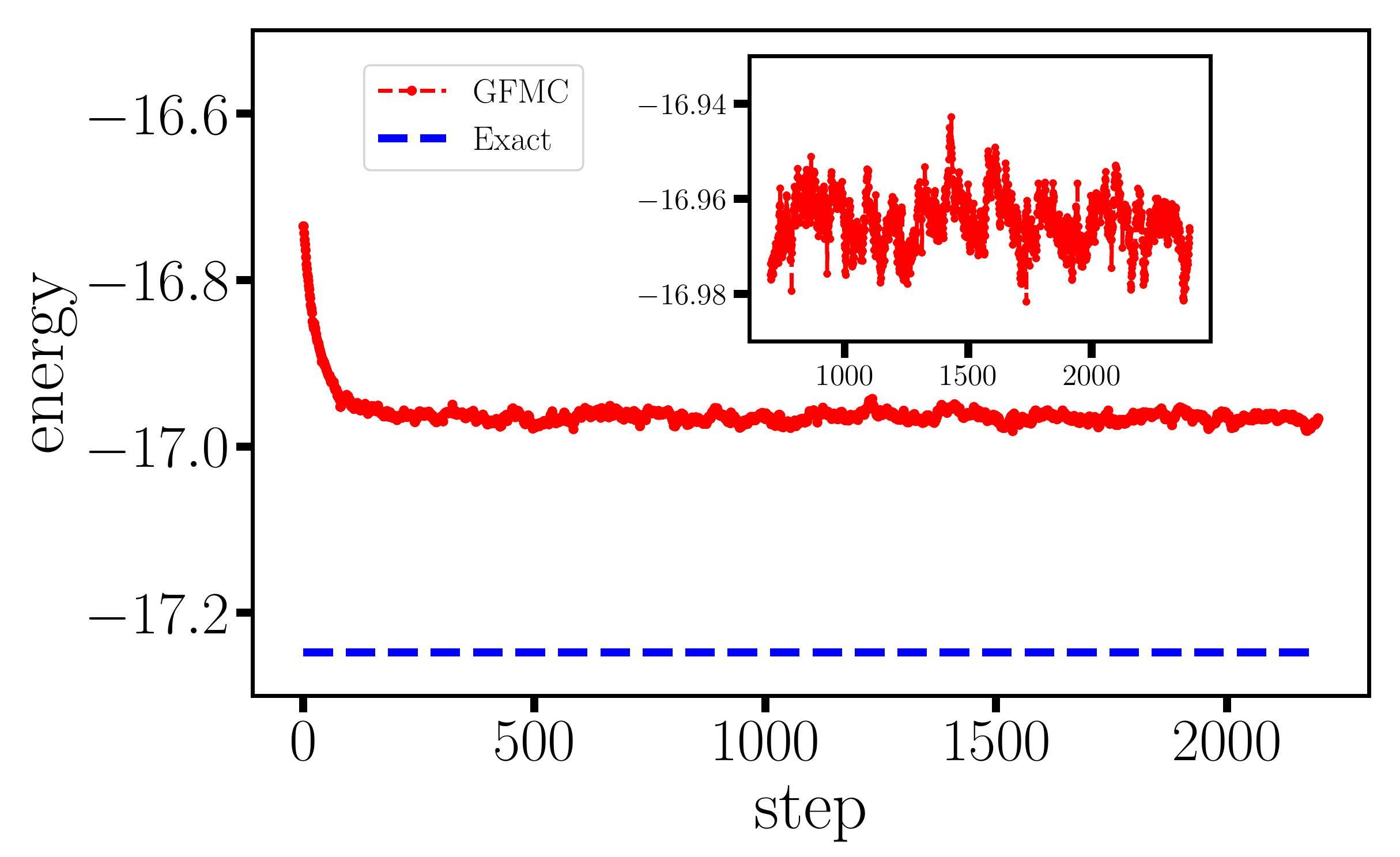}
    	\caption{Similar as Fig.~{\ref{4_4_j2_0}}. Energy of a $6 \times 6$ system with OBC and $J_2 = 0.5$.
    		The guiding wave function is a $D = 4$ PEPS with variational energy $-16.763(4)$ (the energy at step $0$).
    		GFMC energy is $-16.965(1)$ and exact energy (the blue dashed line) for this system is $-17.24733$. 
    	}
    	\label{6_6_j2_05}
    \end{figure}


{\em Green's function Monte Carlo --} In Green's function Monte Carlo \cite{PhysRevB.40.846,PhysRevB.40.2737,PhysRevB.41.4552,PhysRevB.45.7229}, the ground state of a system is also obtained by imaginary time projection
 $|\psi_{g}\rangle \propto\lim_{\beta\rightarrow\infty}\exp(-\beta H)|\psi_{0}\rangle$. The projection length, $\beta$, is then divided into small slices with $\beta = M\tau$ and $\tau$  a small number.  In GFMC, we take the first order expansion of the exponential
function: $\exp(-\tau H)\propto C-H$ with $C$ (or $1/\tau$) a positive real number large enough to ensure all the diagonal elements of $H$ are positive. Then the ground state can formally be
written as
 $|\psi_{g}\rangle\propto\lim_{k\rightarrow\infty}K^{k}|\psi_{0}\rangle$ with $K=C - H$. Mixed estimate is employed in GFMC to calculate the ground state energy as
$E=\frac{\langle\psi_{g}|H|\psi_{G}\rangle}{\langle\psi_{g}|\psi_{G}\rangle}$, which gives the exact ground state energy if $|\psi_g\rangle$ is the
true ground state of $H$. Here we introduce a guiding wave function $\psi_{G}$ whose effect will be discussed late.
We use $S$ to denote spin configuration of the whole system, i.e., $S=(s_{1}^{z},s_{2}^{z},\dots,s_{N}^{z})$,
which is also walker in the sampling process.
We define the kernel as $\widetilde{K}(S^{\prime},S)=\psi_{G}^{*}(S^{\prime})K(S^{\prime},S)/\psi_{G}^{*}(S)$
with $K(S^{\prime},S) =\langle S^{\prime}|K|S\rangle$
and $\psi_{G}(S)  =\langle S|\psi_{G}\rangle$.
Then the ground state energy from mixed estimate is
\begin{eqnarray}
E =\frac{\langle\psi_{g}|H|\psi_{G}\rangle}{\langle\psi_{g}|\psi_{G}\rangle}
 =\frac{\sum_{\{S\}}E_{loc}(S)\psi_{G}(S)\psi_{g}^{*}(S)}{\sum_{\{S\}}\psi_{G}(S)\psi_{g}^{*}(S)}
\label{eq:8}
\end{eqnarray}
where the local energy is defined as 
$
E_{loc}(S)=\frac{\langle S|H|\psi_{G}\rangle}{\langle S|\psi_{G}\rangle}.
$
By introducing $f^{*}(S) =\psi_{G}^{*}(S)\psi_{g}(S)$, we have
\begin{equation}
E=\frac{\sum_{\{S\}}E_{loc}(S)f(S)}{\sum_{\{S\}}f(S)}\label{eq:E_f}
\end{equation}
So we can view $f(S)$ as probability density (if $f(S)$ is non-negative) and take advantage of Monte Carlo techniques, metropolis for example, to evaluate the summation in Eq.~(\ref{eq:E_f}).
It is easy to show
$
f^{*}(S)=\sum \widetilde{K}(S,S_{M})\widetilde{K}(S_{M},S_{M-1})\dots\widetilde{K}(S_{2},S_{1})
 \psi_{0}(S_{1})\psi_{G}^{*}(S_{1})
$ where the summation is over $\{S_{1},S_{2},\dots,S_{M}\}$.

The procedure of GFMC can be summarized as follows. At the first step, we sample $S_{1}$ according to $\psi_{0}(S_{1})\psi_{G}^{*}(S_{1})$,
and set the weight of each walker to $1$. This gives us an ensemble of walkers $\{S_{1}^{i},\omega_{1}^{i}=1\}$.
We usually choose $\psi_0  = \psi_G$ and sample $\{S_1\}$ with probability $|\psi_{G}(S_{1})|^2$.
   Then each walker $\{S_{1}^{i},\omega_{1}^{i}\}$ is propagated to $\{S_{2}^{i},\omega_{2}^{i}\}$
with probability 
$p(S_{2}^{i})=\widetilde{K}(S_{2}^{i},S_{1}^{i})/\beta_{2,1}^{i}$,
where the normalization factor is $\beta_{2,1}^{i}=\sum_{S_{2}^{i}}\widetilde{K}(S_{2}^{i},S_{1}^{i})$. 
After a new walker is chosen, we update the weight of it by multiplying the normalization factor as $\omega_{2}^{i}=\beta_{2,1}^{i}\omega_{1}^{i}$.
This process is repeated and we can start the measurement of energy after equilibrium is reached.
 
 \begin{figure}[t]
 	\includegraphics[width=80mm]{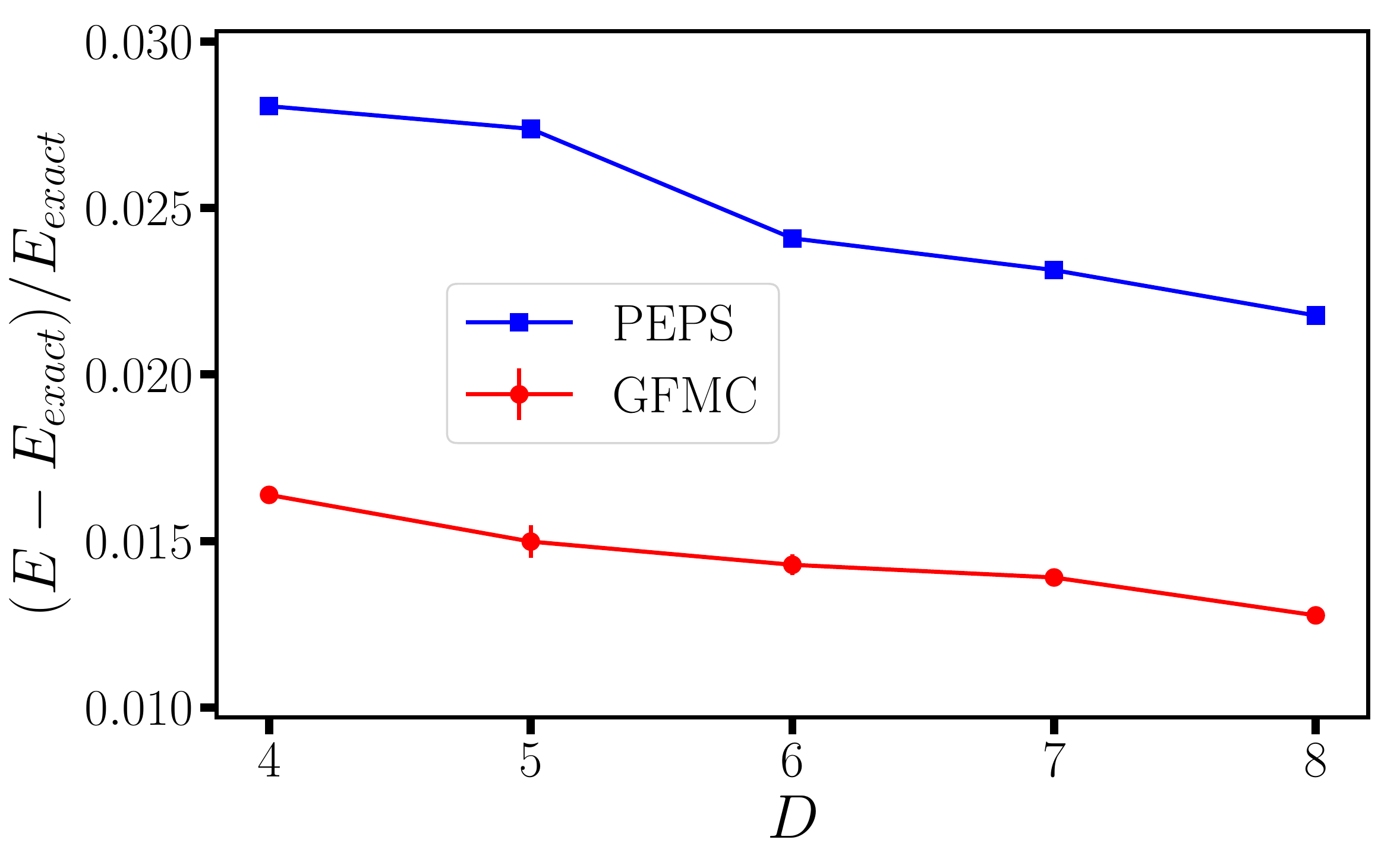}
 	\caption{Comparison of GFMC energy and the energy of corresponding PEPS guiding wave functions for a range of bond dimensions. The relative error
 		of energy is shown. The system size is $6 \times 6$ and $J_2 = 0.5$.}
 	\label{GFMC_6_6}
 \end{figure}
 
When the off-diagonal elements of the $H$ are all non-positive, the ground state of $H$ can be chosen to be
non-negative according to the Perron--Frobenius theorem. Under this circumstance, $f(S)$
is non-negative if we choose
an arbitrary non-negative guiding wave function because it is easily to prove the kernel $K(S^{\prime},S)$ is non-negative. Then we can view $f(S)$ as probability
density without suffering from the negative sign problem.

For Hamiltonian whose off-diagonal elements are not all negative, e.g., when the system is frustrated, we can't ensure the non-negativeness of $f(S)$
and the negative sign problem emerges. Applying the fixed node approximation \cite{PhysRevB.51.13039} can solve this issue. With fixed node approximation, we
actually study an effective
Hamiltonian $H_{eff}$ instead of the original Hamiltonian $H$. The off-diagonal of $H_{eff}$ is defined as $\langle S^{\prime}|H_{eff}|S\rangle=\langle S^{\prime}|H|S\rangle$
if $\psi_G(S^{\prime})H(S^{\prime},S)/\psi_{G}^{*}(S)<0$ and 
it is $0$ if $\psi_G(S^{\prime})H(S^{\prime},S)/\psi_{G}^{*}(S)\ge 0$, which means the off-diagonal elements causing the sign problem are discarded in $H_{eff}$.
The diagonal part is defined as
$\langle S|H_{eff}|S\rangle=\langle S|H|S\rangle+\langle S|V_{sf}|S\rangle$ with $\langle S|V_{sf}|S\rangle=\sum_{S^{\prime}}\langle S|H|S^{\prime}\rangle\frac{\psi_{G}(S^{\prime})}{\psi_{G}(S)}$ where the summation is over all neighboring configurations $S^{\prime}$
of $S$ for which $\psi_G(S^{\prime})H(S^{\prime},S)/\psi_{G}^{*}(S) > 0$. The effect of $V_{sf}$ is a repulsion suppressing the
wave function close to the node which is essential for the energy to be variational \cite{PhysRevB.51.13039}. 
Different from continuum system, both the sign and magnitude of guiding
wave function
effect the accuracy in GFMC for lattice systems \cite{PhysRevB.51.13039}.
The off-diagonal elements of $H_{eff}$ are now all non-positive by definition which allows to obtain
the ground state of it with GFMC without suffering from negative sign problem. It can be proven
GFMC is variational \cite{PhysRevB.51.13039}.
and $E_{eff} \le \langle\psi_{G}|H|\psi_{G}\rangle / \langle\psi_{G}|\psi_{G}\rangle$ which ensures GFMC gives a more accurate 
energy than the variational energy of $\psi_{G}$ \cite{PhysRevB.51.13039}. 
In practice,
a fixed number of walkers are carried in the projection process \cite{PhysRevB.57.11446} and a reconfiguration process is performed periodically 
\cite{PhysRevB.57.11446} to reduce the fluctuation among walkers.  

We can see that $\psi_{G}$ serves as an important
function in the GFMC sampling process when there is no sign problem. When the sign problem is present, $\psi_{G}$ is also used to control the sign problem.
So the quality of $\psi_{G}$ controls both the accuracy and efficiency of GFMC. PEPS are known to be accurate
wave function for $2D$ systems, which makes them good candidates for $\psi_{G}$. In GFMC, we only need to calculate the
overlap between $\psi_{G}$ and walker which is 
is a tensor network also with bond dimension $D$, while in the calculation
the physical quantities of PEPS, contraction of double-layer tensor network with bond dimension $D^2$ is needed. Although it is known that
the rigorous contraction of tensor network in two dimension is fundamentally difficult \cite{PhysRevLett.98.140506}, many effective approximate algorithms exist
\cite{doi:10.1143/JPSJ.65.891,PhysRevLett.99.120601,PhysRevB.80.155131,PhysRevB.81.174411,PhysRevB.90.064425,PhysRevB.86.045139,PhysRevLett.115.180405,PhysRevLett.118.110504,PhysRevB.98.085155} in the literature.

This work is not the first time TNS and GFMC are combined. Many attempts have been made in the past to either
optimize tensor network states \cite{PhysRevLett.99.220602,PhysRevLett.100.040501,PhysRevB.83.134421,PhysRevB.91.165113,PhysRevB.95.195154,PhysRevB.96.085103} with Monte Carlo techniques,
or to take MPS as guiding \cite{PhysRevB.62.14844} or trial wave function \cite{PhysRevB.90.045104} to control the negative sign problem.  
The advance in our work is that we take true 2D TNS, PEPS, as guiding wave function in GFMC, which can reduce the cost of PEPS substantially
and at the same time improve the accuracy over PEPS.   
  
{\em Results --} It is known that the Heisenberg model without $J_2$ term is sign problem free \cite{PhysRevB.40.846,PhysRevB.40.2737,PhysRevB.41.4552,PhysRevB.45.7229}. 
By a rotation of the spin along z-axis on one sub-lattice, sign of the coupling for $x$ and $y$ components is flipped,
and the off-diagonal elements of $H$ in Eq.~(\ref{Ham}) are all negative.
This is the so-called Marshall sign in the ground state of Heisenberg model \cite{doi:10.1098/rspa.1955.0200}
on bipartite lattices.
In the following, we show benchmark results for both $J_2 = 0$ and $J_2 \ne 0$ for $4 \times 4$ and $6 \times 6$ lattice sizes. The ``exact" ground state energies are
from DMRG with truncation error below $10^{-8}$ and with an extrapolation to zero truncation error.
  
In Fig.~{\ref{4_4_j2_0}} we show results for $J_2 = 0$ on a $4 \times 4$ lattice with open boundary conditions.
We first obtain a $D = 4$ PEPS with cluster update, which gives $E_{D=4} = -9.034333$. We then carry out a GFMC calculation
with this PEPS as guiding wave function. As we can see from Fig.~{\ref{4_4_j2_0}}, the final converged energy ($-9.189(1)$) match the exact energy ($-9.189207$)
within error bar, which is reasonable because there is no negative sign problem here. When there is no sign problem, the guiding PEPS wave function plays the role
of important function whose quality only effect the sampling efficiency in GFMC.

We then move to the more challenging $J_2 \ne 0$ case, where the negative sign problem emerges and we need to employ the fixed node approximation. In Fig.~{\ref{6_6_j2_05}}, we show 
the GFMC results for a system on $6 \times 6$ lattice with OBC and $J_2 = 0.5$,
which represents the most difficulty region of the $J_1$-$J_2$ Heisenberg model. A PEPS with $D = 4$ from cluster update is taken as guiding wave function in GFMC.
The variational energy for the $D = 4$ PEPS is $-16.763(4)$ while the energy from GFMC is $-16.965(1)$. This means the error to exact energy ($-17.24733$) is reduced nearly a half
with GFMC
in this case. In Fig.~{\ref{GFMC_6_6}} we show the comparison of converged GFMC energy and the energy of the corresponding PEPS guiding wave function for a range of 
bond dimensions. We can see for all the $D$ values, the GFMC energies are lower than PEPS energy and the errors are reduced by about $40\%$. We want to
emphasize that $J_2 = 0.5$ is the most difficulty region for the $J_1$-$J_2$ Heisenberg model. We anticipate that the improvement in other regions will be even larger (the $J_2 = 0$ result above is an example).

Usually in GFMC calculation, a variational Monte Carlo (VMC) calculation is performed at first to obtain the guiding wave function \cite{PhysRevB.45.7229}. TNS can be also viewed as variational wave functions. Nevertheless, the form of wave function needs to be specified in VMC, while TNS are more general and are systematically
improvable with the increase of bond dimension $D$ \cite{PhysRevLett.118.137202}. Moreover, PEPS satisfy the area law \cite{2004cond.mat..7066V} by design.             
   
{\em Summary and perspectives --} In conclusion, we proposed a new approach to study the ground state properties of quantum many-body systems in which TNS and GFMC are combined to take advantage of both of them. 
Benchmark results for $J_1$-$J_2$ Heisenberg
model on square lattice were provided to demonstrate the effectiveness of this method.
One benefit to combine PEPS and GFMC is that only contraction of single-layer tensor network is needed which reduces the computational complexity substantially (from $D^{12}$ to $D^6$) and
enable the reach of large bond dimension in PEPS, if the optimization process doesn't involve contraction of double-layer tensor network. Nevertheless, after obtaining the
optimized PEPS with full update, we can further improve the energy by taking it as guiding wave function, though the bottleneck is the optimization of PEPS itself in 
full update. 
The energy obtained in GFMC is guaranteed to be variational and lower than the PEPS guiding wave function. 
 Our method can be improved in many aspects. With the stochastic
reconfiguration technique \cite{PhysRevB.61.2599}, the bias from guiding wave function can be reduced. 
We can generalize single PEPS guiding function to a linear combination of PEPS \cite{Huang_2018} which can give lower variational energy.
Generalization to fermionic systems is straightforward \cite{PhysRevB.44.9410,PhysRevLett.72.2442}. We only calculated ground state energy in
this work, but excitation gaps can be calculated easily by enforcing symmetry, conservation of $S_z^{tot}$ for example,
in PEPS. Quantities other than energy can be calculated with the forward propagation technique \cite{PhysRevB.45.7229} without increasing
of computational complexity. Tensor Network States other than PEPS \cite{PhysRevA.74.022320,PhysRevLett.101.110501,PhysRevX.4.011025} can also be adopted as guiding wave function. 
 We believe that this new approach will provide us an accurate and efficient tool in the study of
strongly correlated many-body systems in the future. 

\begin{acknowledgments}
This work is supported by a start-up fund from School of Physics and Astronomy in Shanghai Jiao Tong University. 
We thank Shiwei Zhang for previous discussions about the combination of TNS and QMC.
We also acknowledge the support of computational resource by Shiwei Zhang at the Flatiron Institute.
\end{acknowledgments}

\bibliography{Green_PEPS.bib}

\begin{thebibliography}{100}
\expandafter\ifx\csname natexlab\endcsname\relax\def\natexlab#1{#1}\fi
\expandafter\ifx\csname bibnamefont\endcsname\relax
  \def\bibnamefont#1{#1}\fi
\expandafter\ifx\csname bibfnamefont\endcsname\relax
  \def\bibfnamefont#1{#1}\fi
\expandafter\ifx\csname citenamefont\endcsname\relax
  \def\citenamefont#1{#1}\fi
\expandafter\ifx\csname url\endcsname\relax
  \def\url#1{\texttt{#1}}\fi
\expandafter\ifx\csname urlprefix\endcsname\relax\def\urlprefix{URL }\fi
\providecommand{\bibinfo}[2]{#2}
\providecommand{\eprint}[2][]{\url{#2}}

\bibitem[{\citenamefont{Bednorz and M{\"u}ller}(1986)}]{Bednorz1986}
\bibinfo{author}{\bibfnamefont{J.~G.} \bibnamefont{Bednorz}} \bibnamefont{and}
  \bibinfo{author}{\bibfnamefont{K.~A.} \bibnamefont{M{\"u}ller}},
  \bibinfo{journal}{Zeitschrift f{\"u}r Physik B Condensed Matter}
  \textbf{\bibinfo{volume}{64}}, \bibinfo{pages}{189} (\bibinfo{year}{1986}),
  ISSN \bibinfo{issn}{1431-584X},
  \urlprefix\url{https://doi.org/10.1007/BF01303701}.

\bibitem[{\citenamefont{Tsui et~al.}(1982)\citenamefont{Tsui, Stormer, and
  Gossard}}]{PhysRevLett.48.1559}
\bibinfo{author}{\bibfnamefont{D.~C.} \bibnamefont{Tsui}},
  \bibinfo{author}{\bibfnamefont{H.~L.} \bibnamefont{Stormer}},
  \bibnamefont{and} \bibinfo{author}{\bibfnamefont{A.~C.}
  \bibnamefont{Gossard}}, \bibinfo{journal}{Phys. Rev. Lett.}
  \textbf{\bibinfo{volume}{48}}, \bibinfo{pages}{1559} (\bibinfo{year}{1982}),
  \urlprefix\url{https://link.aps.org/doi/10.1103/PhysRevLett.48.1559}.

\bibitem[{\citenamefont{Laughlin}(1983)}]{PhysRevLett.50.1395}
\bibinfo{author}{\bibfnamefont{R.~B.} \bibnamefont{Laughlin}},
  \bibinfo{journal}{Phys. Rev. Lett.} \textbf{\bibinfo{volume}{50}},
  \bibinfo{pages}{1395} (\bibinfo{year}{1983}),
  \urlprefix\url{https://link.aps.org/doi/10.1103/PhysRevLett.50.1395}.

\bibitem[{\citenamefont{Zhou et~al.}(2017)\citenamefont{Zhou, Kanoda, and
  Ng}}]{RevModPhys.89.025003}
\bibinfo{author}{\bibfnamefont{Y.}~\bibnamefont{Zhou}},
  \bibinfo{author}{\bibfnamefont{K.}~\bibnamefont{Kanoda}}, \bibnamefont{and}
  \bibinfo{author}{\bibfnamefont{T.-K.} \bibnamefont{Ng}},
  \bibinfo{journal}{Rev. Mod. Phys.} \textbf{\bibinfo{volume}{89}},
  \bibinfo{pages}{025003} (\bibinfo{year}{2017}),
  \urlprefix\url{https://link.aps.org/doi/10.1103/RevModPhys.89.025003}.

\bibitem[{\citenamefont{Lieb and Wu}(1968)}]{PhysRevLett.20.1445}
\bibinfo{author}{\bibfnamefont{E.~H.} \bibnamefont{Lieb}} \bibnamefont{and}
  \bibinfo{author}{\bibfnamefont{F.~Y.} \bibnamefont{Wu}},
  \bibinfo{journal}{Phys. Rev. Lett.} \textbf{\bibinfo{volume}{20}},
  \bibinfo{pages}{1445} (\bibinfo{year}{1968}),
  \urlprefix\url{https://link.aps.org/doi/10.1103/PhysRevLett.20.1445}.

\bibitem[{\citenamefont{White}(1992)}]{PhysRevLett.69.2863}
\bibinfo{author}{\bibfnamefont{S.~R.} \bibnamefont{White}},
  \bibinfo{journal}{Phys. Rev. Lett.} \textbf{\bibinfo{volume}{69}},
  \bibinfo{pages}{2863} (\bibinfo{year}{1992}),
  \urlprefix\url{https://link.aps.org/doi/10.1103/PhysRevLett.69.2863}.

\bibitem[{\citenamefont{White}(1993)}]{PhysRevB.48.10345}
\bibinfo{author}{\bibfnamefont{S.~R.} \bibnamefont{White}},
  \bibinfo{journal}{Phys. Rev. B} \textbf{\bibinfo{volume}{48}},
  \bibinfo{pages}{10345} (\bibinfo{year}{1993}),
  \urlprefix\url{https://link.aps.org/doi/10.1103/PhysRevB.48.10345}.

\bibitem[{\citenamefont{Schollw\"ock}(2005)}]{RevModPhys.77.259}
\bibinfo{author}{\bibfnamefont{U.}~\bibnamefont{Schollw\"ock}},
  \bibinfo{journal}{Rev. Mod. Phys.} \textbf{\bibinfo{volume}{77}},
  \bibinfo{pages}{259} (\bibinfo{year}{2005}),
  \urlprefix\url{https://link.aps.org/doi/10.1103/RevModPhys.77.259}.

\bibitem[{\citenamefont{Schollw\"ock}(2011)}]{SCHOLLWOCK201196}
\bibinfo{author}{\bibfnamefont{U.}~\bibnamefont{Schollw\"ock}},
  \bibinfo{journal}{Annals of Physics} \textbf{\bibinfo{volume}{326}},
  \bibinfo{pages}{96 } (\bibinfo{year}{2011}), ISSN \bibinfo{issn}{0003-4916},
  \bibinfo{note}{january 2011 Special Issue},
  \urlprefix\url{http://www.sciencedirect.com/science/article/pii/S0003491610001752}.

\bibitem[{\citenamefont{\"Ostlund and Rommer}(1995)}]{PhysRevLett.75.3537}
\bibinfo{author}{\bibfnamefont{S.}~\bibnamefont{\"Ostlund}} \bibnamefont{and}
  \bibinfo{author}{\bibfnamefont{S.}~\bibnamefont{Rommer}},
  \bibinfo{journal}{Phys. Rev. Lett.} \textbf{\bibinfo{volume}{75}},
  \bibinfo{pages}{3537} (\bibinfo{year}{1995}),
  \urlprefix\url{https://link.aps.org/doi/10.1103/PhysRevLett.75.3537}.

\bibitem[{\citenamefont{Affleck et~al.}(1987)\citenamefont{Affleck, Kennedy,
  Lieb, and Tasaki}}]{PhysRevLett.59.799}
\bibinfo{author}{\bibfnamefont{I.}~\bibnamefont{Affleck}},
  \bibinfo{author}{\bibfnamefont{T.}~\bibnamefont{Kennedy}},
  \bibinfo{author}{\bibfnamefont{E.~H.} \bibnamefont{Lieb}}, \bibnamefont{and}
  \bibinfo{author}{\bibfnamefont{H.}~\bibnamefont{Tasaki}},
  \bibinfo{journal}{Phys. Rev. Lett.} \textbf{\bibinfo{volume}{59}},
  \bibinfo{pages}{799} (\bibinfo{year}{1987}),
  \urlprefix\url{https://link.aps.org/doi/10.1103/PhysRevLett.59.799}.

\bibitem[{\citenamefont{Baxter}(1968)}]{doi:10.1063/1.1664623}
\bibinfo{author}{\bibfnamefont{R.~J.} \bibnamefont{Baxter}},
  \bibinfo{journal}{Journal of Mathematical Physics}
  \textbf{\bibinfo{volume}{9}}, \bibinfo{pages}{650} (\bibinfo{year}{1968}),
  \eprint{https://doi.org/10.1063/1.1664623},
  \urlprefix\url{https://doi.org/10.1063/1.1664623}.

\bibitem[{\citenamefont{Eisert et~al.}(2010)\citenamefont{Eisert, Cramer, and
  Plenio}}]{RevModPhys.82.277}
\bibinfo{author}{\bibfnamefont{J.}~\bibnamefont{Eisert}},
  \bibinfo{author}{\bibfnamefont{M.}~\bibnamefont{Cramer}}, \bibnamefont{and}
  \bibinfo{author}{\bibfnamefont{M.~B.} \bibnamefont{Plenio}},
  \bibinfo{journal}{Rev. Mod. Phys.} \textbf{\bibinfo{volume}{82}},
  \bibinfo{pages}{277} (\bibinfo{year}{2010}),
  \urlprefix\url{https://link.aps.org/doi/10.1103/RevModPhys.82.277}.

\bibitem[{\citenamefont{Verstraete et~al.}(2008)\citenamefont{Verstraete, Murg,
  and Cirac}}]{doi:10.1080/14789940801912366}
\bibinfo{author}{\bibfnamefont{F.}~\bibnamefont{Verstraete}},
  \bibinfo{author}{\bibfnamefont{V.}~\bibnamefont{Murg}}, \bibnamefont{and}
  \bibinfo{author}{\bibfnamefont{J.}~\bibnamefont{Cirac}},
  \bibinfo{journal}{Advances in Physics} \textbf{\bibinfo{volume}{57}},
  \bibinfo{pages}{143} (\bibinfo{year}{2008}),
  \eprint{https://doi.org/10.1080/14789940801912366},
  \urlprefix\url{https://doi.org/10.1080/14789940801912366}.

\bibitem[{\citenamefont{Or\'us}(2014)}]{ORUS2014117}
\bibinfo{author}{\bibfnamefont{R.}~\bibnamefont{Or\'us}},
  \bibinfo{journal}{Annals of Physics} \textbf{\bibinfo{volume}{349}},
  \bibinfo{pages}{117 } (\bibinfo{year}{2014}), ISSN \bibinfo{issn}{0003-4916},
  \urlprefix\url{http://www.sciencedirect.com/science/article/pii/S0003491614001596}.

\bibitem[{\citenamefont{Chen et~al.}(2011)\citenamefont{Chen, Gu, and
  Wen}}]{PhysRevB.83.035107}
\bibinfo{author}{\bibfnamefont{X.}~\bibnamefont{Chen}},
  \bibinfo{author}{\bibfnamefont{Z.-C.} \bibnamefont{Gu}}, \bibnamefont{and}
  \bibinfo{author}{\bibfnamefont{X.-G.} \bibnamefont{Wen}},
  \bibinfo{journal}{Phys. Rev. B} \textbf{\bibinfo{volume}{83}},
  \bibinfo{pages}{035107} (\bibinfo{year}{2011}),
  \urlprefix\url{https://link.aps.org/doi/10.1103/PhysRevB.83.035107}.

\bibitem[{\citenamefont{Schuch et~al.}(2011)\citenamefont{Schuch,
  P\'erez-Garc\'{\i}a, and Cirac}}]{PhysRevB.84.165139}
\bibinfo{author}{\bibfnamefont{N.}~\bibnamefont{Schuch}},
  \bibinfo{author}{\bibfnamefont{D.}~\bibnamefont{P\'erez-Garc\'{\i}a}},
  \bibnamefont{and} \bibinfo{author}{\bibfnamefont{I.}~\bibnamefont{Cirac}},
  \bibinfo{journal}{Phys. Rev. B} \textbf{\bibinfo{volume}{84}},
  \bibinfo{pages}{165139} (\bibinfo{year}{2011}),
  \urlprefix\url{https://link.aps.org/doi/10.1103/PhysRevB.84.165139}.

\bibitem[{\citenamefont{Stoudenmire and
  White}(2012)}]{doi:10.1146/annurev-conmatphys-020911-125018}
\bibinfo{author}{\bibfnamefont{E.}~\bibnamefont{Stoudenmire}} \bibnamefont{and}
  \bibinfo{author}{\bibfnamefont{S.~R.} \bibnamefont{White}},
  \bibinfo{journal}{Annual Review of Condensed Matter Physics}
  \textbf{\bibinfo{volume}{3}}, \bibinfo{pages}{111} (\bibinfo{year}{2012}),
  \eprint{https://doi.org/10.1146/annurev-conmatphys-020911-125018},
  \urlprefix\url{https://doi.org/10.1146/annurev-conmatphys-020911-125018}.

\bibitem[{\citenamefont{Yan et~al.}(2011)\citenamefont{Yan, Huse, and
  White}}]{Yan1173}
\bibinfo{author}{\bibfnamefont{S.}~\bibnamefont{Yan}},
  \bibinfo{author}{\bibfnamefont{D.~A.} \bibnamefont{Huse}}, \bibnamefont{and}
  \bibinfo{author}{\bibfnamefont{S.~R.} \bibnamefont{White}},
  \bibinfo{journal}{Science} \textbf{\bibinfo{volume}{332}},
  \bibinfo{pages}{1173} (\bibinfo{year}{2011}), ISSN \bibinfo{issn}{0036-8075},
  \eprint{https://science.sciencemag.org/content/332/6034/1173.full.pdf},
  \urlprefix\url{https://science.sciencemag.org/content/332/6034/1173}.

\bibitem[{\citenamefont{LeBlanc et~al.}(2015)\citenamefont{LeBlanc, Antipov,
  Becca, Bulik, Chan, Chung, Deng, Ferrero, Henderson, Jim\'enez-Hoyos
  et~al.}}]{PhysRevX.5.041041}
\bibinfo{author}{\bibfnamefont{J.~P.~F.} \bibnamefont{LeBlanc}},
  \bibinfo{author}{\bibfnamefont{A.~E.} \bibnamefont{Antipov}},
  \bibinfo{author}{\bibfnamefont{F.}~\bibnamefont{Becca}},
  \bibinfo{author}{\bibfnamefont{I.~W.} \bibnamefont{Bulik}},
  \bibinfo{author}{\bibfnamefont{G.~K.-L.} \bibnamefont{Chan}},
  \bibinfo{author}{\bibfnamefont{C.-M.} \bibnamefont{Chung}},
  \bibinfo{author}{\bibfnamefont{Y.}~\bibnamefont{Deng}},
  \bibinfo{author}{\bibfnamefont{M.}~\bibnamefont{Ferrero}},
  \bibinfo{author}{\bibfnamefont{T.~M.} \bibnamefont{Henderson}},
  \bibinfo{author}{\bibfnamefont{C.~A.} \bibnamefont{Jim\'enez-Hoyos}},
  \bibnamefont{et~al.} (\bibinfo{collaboration}{Simons Collaboration on the
  Many-Electron Problem}), \bibinfo{journal}{Phys. Rev. X}
  \textbf{\bibinfo{volume}{5}}, \bibinfo{pages}{041041} (\bibinfo{year}{2015}),
  \urlprefix\url{https://link.aps.org/doi/10.1103/PhysRevX.5.041041}.

\bibitem[{\citenamefont{Zheng et~al.}(2017)\citenamefont{Zheng, Chung, Corboz,
  Ehlers, Qin, Noack, Shi, White, Zhang, and Chan}}]{Zheng1155}
\bibinfo{author}{\bibfnamefont{B.-X.} \bibnamefont{Zheng}},
  \bibinfo{author}{\bibfnamefont{C.-M.} \bibnamefont{Chung}},
  \bibinfo{author}{\bibfnamefont{P.}~\bibnamefont{Corboz}},
  \bibinfo{author}{\bibfnamefont{G.}~\bibnamefont{Ehlers}},
  \bibinfo{author}{\bibfnamefont{M.-P.} \bibnamefont{Qin}},
  \bibinfo{author}{\bibfnamefont{R.~M.} \bibnamefont{Noack}},
  \bibinfo{author}{\bibfnamefont{H.}~\bibnamefont{Shi}},
  \bibinfo{author}{\bibfnamefont{S.~R.} \bibnamefont{White}},
  \bibinfo{author}{\bibfnamefont{S.}~\bibnamefont{Zhang}}, \bibnamefont{and}
  \bibinfo{author}{\bibfnamefont{G.~K.-L.} \bibnamefont{Chan}},
  \bibinfo{journal}{Science} \textbf{\bibinfo{volume}{358}},
  \bibinfo{pages}{1155} (\bibinfo{year}{2017}), ISSN \bibinfo{issn}{0036-8075},
  \eprint{http://science.sciencemag.org/content/358/6367/1155.full.pdf},
  \urlprefix\url{http://science.sciencemag.org/content/358/6367/1155}.

\bibitem[{\citenamefont{Qin et~al.}(2020)\citenamefont{Qin, Chung, Shi, Vitali,
  Hubig, Schollw\"ock, White, and Zhang}}]{PhysRevX.10.031016}
\bibinfo{author}{\bibfnamefont{M.}~\bibnamefont{Qin}},
  \bibinfo{author}{\bibfnamefont{C.-M.} \bibnamefont{Chung}},
  \bibinfo{author}{\bibfnamefont{H.}~\bibnamefont{Shi}},
  \bibinfo{author}{\bibfnamefont{E.}~\bibnamefont{Vitali}},
  \bibinfo{author}{\bibfnamefont{C.}~\bibnamefont{Hubig}},
  \bibinfo{author}{\bibfnamefont{U.}~\bibnamefont{Schollw\"ock}},
  \bibinfo{author}{\bibfnamefont{S.~R.} \bibnamefont{White}}, \bibnamefont{and}
  \bibinfo{author}{\bibfnamefont{S.}~\bibnamefont{Zhang}}
  (\bibinfo{collaboration}{Simons Collaboration on the Many-Electron Problem}),
  \bibinfo{journal}{Phys. Rev. X} \textbf{\bibinfo{volume}{10}},
  \bibinfo{pages}{031016} (\bibinfo{year}{2020}),
  \urlprefix\url{https://link.aps.org/doi/10.1103/PhysRevX.10.031016}.

\bibitem[{\citenamefont{Liang and Pang}(1994)}]{PhysRevB.49.9214}
\bibinfo{author}{\bibfnamefont{S.}~\bibnamefont{Liang}} \bibnamefont{and}
  \bibinfo{author}{\bibfnamefont{H.}~\bibnamefont{Pang}},
  \bibinfo{journal}{Phys. Rev. B} \textbf{\bibinfo{volume}{49}},
  \bibinfo{pages}{9214} (\bibinfo{year}{1994}),
  \urlprefix\url{https://link.aps.org/doi/10.1103/PhysRevB.49.9214}.

\bibitem[{\citenamefont{{Verstraete} and {Cirac}}(2004)}]{2004cond.mat..7066V}
\bibinfo{author}{\bibfnamefont{F.}~\bibnamefont{{Verstraete}}}
  \bibnamefont{and} \bibinfo{author}{\bibfnamefont{J.~I.}
  \bibnamefont{{Cirac}}}, \bibinfo{journal}{arXiv e-prints}
  \bibinfo{eid}{cond-mat/0407066} (\bibinfo{year}{2004}),
  \eprint{cond-mat/0407066}.

\bibitem[{\citenamefont{Jiang et~al.}(2008)\citenamefont{Jiang, Weng, and
  Xiang}}]{PhysRevLett.101.090603}
\bibinfo{author}{\bibfnamefont{H.~C.} \bibnamefont{Jiang}},
  \bibinfo{author}{\bibfnamefont{Z.~Y.} \bibnamefont{Weng}}, \bibnamefont{and}
  \bibinfo{author}{\bibfnamefont{T.}~\bibnamefont{Xiang}},
  \bibinfo{journal}{Phys. Rev. Lett.} \textbf{\bibinfo{volume}{101}},
  \bibinfo{pages}{090603} (\bibinfo{year}{2008}),
  \urlprefix\url{https://link.aps.org/doi/10.1103/PhysRevLett.101.090603}.

\bibitem[{\citenamefont{Jordan et~al.}(2008)\citenamefont{Jordan, Or\'us,
  Vidal, Verstraete, and Cirac}}]{PhysRevLett.101.250602}
\bibinfo{author}{\bibfnamefont{J.}~\bibnamefont{Jordan}},
  \bibinfo{author}{\bibfnamefont{R.}~\bibnamefont{Or\'us}},
  \bibinfo{author}{\bibfnamefont{G.}~\bibnamefont{Vidal}},
  \bibinfo{author}{\bibfnamefont{F.}~\bibnamefont{Verstraete}},
  \bibnamefont{and} \bibinfo{author}{\bibfnamefont{J.~I.} \bibnamefont{Cirac}},
  \bibinfo{journal}{Phys. Rev. Lett.} \textbf{\bibinfo{volume}{101}},
  \bibinfo{pages}{250602} (\bibinfo{year}{2008}),
  \urlprefix\url{https://link.aps.org/doi/10.1103/PhysRevLett.101.250602}.

\bibitem[{\citenamefont{Corboz et~al.}(2011)\citenamefont{Corboz, L\"auchli,
  Penc, Troyer, and Mila}}]{PhysRevLett.107.215301}
\bibinfo{author}{\bibfnamefont{P.}~\bibnamefont{Corboz}},
  \bibinfo{author}{\bibfnamefont{A.~M.} \bibnamefont{L\"auchli}},
  \bibinfo{author}{\bibfnamefont{K.}~\bibnamefont{Penc}},
  \bibinfo{author}{\bibfnamefont{M.}~\bibnamefont{Troyer}}, \bibnamefont{and}
  \bibinfo{author}{\bibfnamefont{F.}~\bibnamefont{Mila}},
  \bibinfo{journal}{Phys. Rev. Lett.} \textbf{\bibinfo{volume}{107}},
  \bibinfo{pages}{215301} (\bibinfo{year}{2011}),
  \urlprefix\url{https://link.aps.org/doi/10.1103/PhysRevLett.107.215301}.

\bibitem[{\citenamefont{Corboz et~al.}(2014)\citenamefont{Corboz, Rice, and
  Troyer}}]{PhysRevLett.113.046402}
\bibinfo{author}{\bibfnamefont{P.}~\bibnamefont{Corboz}},
  \bibinfo{author}{\bibfnamefont{T.~M.} \bibnamefont{Rice}}, \bibnamefont{and}
  \bibinfo{author}{\bibfnamefont{M.}~\bibnamefont{Troyer}},
  \bibinfo{journal}{Phys. Rev. Lett.} \textbf{\bibinfo{volume}{113}},
  \bibinfo{pages}{046402} (\bibinfo{year}{2014}),
  \urlprefix\url{https://link.aps.org/doi/10.1103/PhysRevLett.113.046402}.

\bibitem[{\citenamefont{Liao et~al.}(2017)\citenamefont{Liao, Xie, Chen, Liu,
  Xie, Huang, Normand, and Xiang}}]{PhysRevLett.118.137202}
\bibinfo{author}{\bibfnamefont{H.~J.} \bibnamefont{Liao}},
  \bibinfo{author}{\bibfnamefont{Z.~Y.} \bibnamefont{Xie}},
  \bibinfo{author}{\bibfnamefont{J.}~\bibnamefont{Chen}},
  \bibinfo{author}{\bibfnamefont{Z.~Y.} \bibnamefont{Liu}},
  \bibinfo{author}{\bibfnamefont{H.~D.} \bibnamefont{Xie}},
  \bibinfo{author}{\bibfnamefont{R.~Z.} \bibnamefont{Huang}},
  \bibinfo{author}{\bibfnamefont{B.}~\bibnamefont{Normand}}, \bibnamefont{and}
  \bibinfo{author}{\bibfnamefont{T.}~\bibnamefont{Xiang}},
  \bibinfo{journal}{Phys. Rev. Lett.} \textbf{\bibinfo{volume}{118}},
  \bibinfo{pages}{137202} (\bibinfo{year}{2017}),
  \urlprefix\url{https://link.aps.org/doi/10.1103/PhysRevLett.118.137202}.

\bibitem[{\citenamefont{Gioev and Klich}(2006)}]{PhysRevLett.96.100503}
\bibinfo{author}{\bibfnamefont{D.}~\bibnamefont{Gioev}} \bibnamefont{and}
  \bibinfo{author}{\bibfnamefont{I.}~\bibnamefont{Klich}},
  \bibinfo{journal}{Phys. Rev. Lett.} \textbf{\bibinfo{volume}{96}},
  \bibinfo{pages}{100503} (\bibinfo{year}{2006}),
  \urlprefix\url{https://link.aps.org/doi/10.1103/PhysRevLett.96.100503}.

\bibitem[{\citenamefont{Or\'us and Vidal}(2009)}]{PhysRevB.80.094403}
\bibinfo{author}{\bibfnamefont{R.}~\bibnamefont{Or\'us}} \bibnamefont{and}
  \bibinfo{author}{\bibfnamefont{G.}~\bibnamefont{Vidal}},
  \bibinfo{journal}{Phys. Rev. B} \textbf{\bibinfo{volume}{80}},
  \bibinfo{pages}{094403} (\bibinfo{year}{2009}),
  \urlprefix\url{https://link.aps.org/doi/10.1103/PhysRevB.80.094403}.

\bibitem[{\citenamefont{Phien et~al.}(2015)\citenamefont{Phien, Bengua, Tuan,
  Corboz, and Or\'us}}]{PhysRevB.92.035142}
\bibinfo{author}{\bibfnamefont{H.~N.} \bibnamefont{Phien}},
  \bibinfo{author}{\bibfnamefont{J.~A.} \bibnamefont{Bengua}},
  \bibinfo{author}{\bibfnamefont{H.~D.} \bibnamefont{Tuan}},
  \bibinfo{author}{\bibfnamefont{P.}~\bibnamefont{Corboz}}, \bibnamefont{and}
  \bibinfo{author}{\bibfnamefont{R.}~\bibnamefont{Or\'us}},
  \bibinfo{journal}{Phys. Rev. B} \textbf{\bibinfo{volume}{92}},
  \bibinfo{pages}{035142} (\bibinfo{year}{2015}),
  \urlprefix\url{https://link.aps.org/doi/10.1103/PhysRevB.92.035142}.

\bibitem[{\citenamefont{Ran et~al.}(2007)\citenamefont{Ran, Hermele, Lee, and
  Wen}}]{PhysRevLett.98.117205}
\bibinfo{author}{\bibfnamefont{Y.}~\bibnamefont{Ran}},
  \bibinfo{author}{\bibfnamefont{M.}~\bibnamefont{Hermele}},
  \bibinfo{author}{\bibfnamefont{P.~A.} \bibnamefont{Lee}}, \bibnamefont{and}
  \bibinfo{author}{\bibfnamefont{X.-G.} \bibnamefont{Wen}},
  \bibinfo{journal}{Phys. Rev. Lett.} \textbf{\bibinfo{volume}{98}},
  \bibinfo{pages}{117205} (\bibinfo{year}{2007}),
  \urlprefix\url{https://link.aps.org/doi/10.1103/PhysRevLett.98.117205}.

\bibitem[{\citenamefont{Depenbrock et~al.}(2012)\citenamefont{Depenbrock,
  McCulloch, and Schollw\"ock}}]{PhysRevLett.109.067201}
\bibinfo{author}{\bibfnamefont{S.}~\bibnamefont{Depenbrock}},
  \bibinfo{author}{\bibfnamefont{I.~P.} \bibnamefont{McCulloch}},
  \bibnamefont{and}
  \bibinfo{author}{\bibfnamefont{U.}~\bibnamefont{Schollw\"ock}},
  \bibinfo{journal}{Phys. Rev. Lett.} \textbf{\bibinfo{volume}{109}},
  \bibinfo{pages}{067201} (\bibinfo{year}{2012}),
  \urlprefix\url{https://link.aps.org/doi/10.1103/PhysRevLett.109.067201}.

\bibitem[{\citenamefont{Iqbal et~al.}(2013)\citenamefont{Iqbal, Becca, Sorella,
  and Poilblanc}}]{PhysRevB.87.060405}
\bibinfo{author}{\bibfnamefont{Y.}~\bibnamefont{Iqbal}},
  \bibinfo{author}{\bibfnamefont{F.}~\bibnamefont{Becca}},
  \bibinfo{author}{\bibfnamefont{S.}~\bibnamefont{Sorella}}, \bibnamefont{and}
  \bibinfo{author}{\bibfnamefont{D.}~\bibnamefont{Poilblanc}},
  \bibinfo{journal}{Phys. Rev. B} \textbf{\bibinfo{volume}{87}},
  \bibinfo{pages}{060405} (\bibinfo{year}{2013}),
  \urlprefix\url{https://link.aps.org/doi/10.1103/PhysRevB.87.060405}.

\bibitem[{\citenamefont{Mei et~al.}(2017)\citenamefont{Mei, Chen, He, and
  Wen}}]{PhysRevB.95.235107}
\bibinfo{author}{\bibfnamefont{J.-W.} \bibnamefont{Mei}},
  \bibinfo{author}{\bibfnamefont{J.-Y.} \bibnamefont{Chen}},
  \bibinfo{author}{\bibfnamefont{H.}~\bibnamefont{He}}, \bibnamefont{and}
  \bibinfo{author}{\bibfnamefont{X.-G.} \bibnamefont{Wen}},
  \bibinfo{journal}{Phys. Rev. B} \textbf{\bibinfo{volume}{95}},
  \bibinfo{pages}{235107} (\bibinfo{year}{2017}),
  \urlprefix\url{https://link.aps.org/doi/10.1103/PhysRevB.95.235107}.

\bibitem[{\citenamefont{Blankenbecler et~al.}(1981)\citenamefont{Blankenbecler,
  Scalapino, and Sugar}}]{PhysRevD.24.2278}
\bibinfo{author}{\bibfnamefont{R.}~\bibnamefont{Blankenbecler}},
  \bibinfo{author}{\bibfnamefont{D.~J.} \bibnamefont{Scalapino}},
  \bibnamefont{and} \bibinfo{author}{\bibfnamefont{R.~L.} \bibnamefont{Sugar}},
  \bibinfo{journal}{Phys. Rev. D} \textbf{\bibinfo{volume}{24}},
  \bibinfo{pages}{2278} (\bibinfo{year}{1981}),
  \urlprefix\url{https://link.aps.org/doi/10.1103/PhysRevD.24.2278}.

\bibitem[{\citenamefont{Assaad and Evertz}(2008)}]{Assaad2008}
\bibinfo{author}{\bibfnamefont{F.}~\bibnamefont{Assaad}} \bibnamefont{and}
  \bibinfo{author}{\bibfnamefont{H.}~\bibnamefont{Evertz}}, pp.
  \bibinfo{pages}{277--356} (\bibinfo{year}{2008}),
  \urlprefix\url{https://doi.org/10.1007/978-3-540-74686-7_10}.

\bibitem[{\citenamefont{Zhang}(2004)}]{Zhang2004}
\bibinfo{author}{\bibfnamefont{S.}~\bibnamefont{Zhang}},
  \emph{\bibinfo{title}{Quantum Monte Carlo Methods for Strongly Correlated
  Electron Systems}} (\bibinfo{publisher}{Springer New York},
  \bibinfo{address}{New York, NY}, \bibinfo{year}{2004}), pp.
  \bibinfo{pages}{39--74}, ISBN \bibinfo{isbn}{978-0-387-21717-8},
  \urlprefix\url{https://doi.org/10.1007/0-387-21717-7_2}.

\bibitem[{\citenamefont{Loh et~al.}(1990)\citenamefont{Loh, Gubernatis,
  Scalettar, White, Scalapino, and Sugar}}]{PhysRevB.41.9301}
\bibinfo{author}{\bibfnamefont{E.~Y.} \bibnamefont{Loh}},
  \bibinfo{author}{\bibfnamefont{J.~E.} \bibnamefont{Gubernatis}},
  \bibinfo{author}{\bibfnamefont{R.~T.} \bibnamefont{Scalettar}},
  \bibinfo{author}{\bibfnamefont{S.~R.} \bibnamefont{White}},
  \bibinfo{author}{\bibfnamefont{D.~J.} \bibnamefont{Scalapino}},
  \bibnamefont{and} \bibinfo{author}{\bibfnamefont{R.~L.} \bibnamefont{Sugar}},
  \bibinfo{journal}{Phys. Rev. B} \textbf{\bibinfo{volume}{41}},
  \bibinfo{pages}{9301} (\bibinfo{year}{1990}),
  \urlprefix\url{https://link.aps.org/doi/10.1103/PhysRevB.41.9301}.

\bibitem[{\citenamefont{Troyer and Wiese}(2005)}]{PhysRevLett.94.170201}
\bibinfo{author}{\bibfnamefont{M.}~\bibnamefont{Troyer}} \bibnamefont{and}
  \bibinfo{author}{\bibfnamefont{U.-J.} \bibnamefont{Wiese}},
  \bibinfo{journal}{Phys. Rev. Lett.} \textbf{\bibinfo{volume}{94}},
  \bibinfo{pages}{170201} (\bibinfo{year}{2005}),
  \urlprefix\url{https://link.aps.org/doi/10.1103/PhysRevLett.94.170201}.

\bibitem[{\citenamefont{Kalos}(1962)}]{PhysRev.128.1791}
\bibinfo{author}{\bibfnamefont{M.~H.} \bibnamefont{Kalos}},
  \bibinfo{journal}{Phys. Rev.} \textbf{\bibinfo{volume}{128}},
  \bibinfo{pages}{1791} (\bibinfo{year}{1962}),
  \urlprefix\url{https://link.aps.org/doi/10.1103/PhysRev.128.1791}.

\bibitem[{\citenamefont{Ceperley and Alder}(1980)}]{PhysRevLett.45.566}
\bibinfo{author}{\bibfnamefont{D.~M.} \bibnamefont{Ceperley}} \bibnamefont{and}
  \bibinfo{author}{\bibfnamefont{B.~J.} \bibnamefont{Alder}},
  \bibinfo{journal}{Phys. Rev. Lett.} \textbf{\bibinfo{volume}{45}},
  \bibinfo{pages}{566} (\bibinfo{year}{1980}),
  \urlprefix\url{https://link.aps.org/doi/10.1103/PhysRevLett.45.566}.

\bibitem[{\citenamefont{Foulkes et~al.}(2001)\citenamefont{Foulkes, Mitas,
  Needs, and Rajagopal}}]{RevModPhys.73.33}
\bibinfo{author}{\bibfnamefont{W.~M.~C.} \bibnamefont{Foulkes}},
  \bibinfo{author}{\bibfnamefont{L.}~\bibnamefont{Mitas}},
  \bibinfo{author}{\bibfnamefont{R.~J.} \bibnamefont{Needs}}, \bibnamefont{and}
  \bibinfo{author}{\bibfnamefont{G.}~\bibnamefont{Rajagopal}},
  \bibinfo{journal}{Rev. Mod. Phys.} \textbf{\bibinfo{volume}{73}},
  \bibinfo{pages}{33} (\bibinfo{year}{2001}),
  \urlprefix\url{https://link.aps.org/doi/10.1103/RevModPhys.73.33}.

\bibitem[{\citenamefont{Zhang et~al.}(1997)\citenamefont{Zhang, Carlson, and
  Gubernatis}}]{PhysRevB.55.7464}
\bibinfo{author}{\bibfnamefont{S.}~\bibnamefont{Zhang}},
  \bibinfo{author}{\bibfnamefont{J.}~\bibnamefont{Carlson}}, \bibnamefont{and}
  \bibinfo{author}{\bibfnamefont{J.~E.} \bibnamefont{Gubernatis}},
  \bibinfo{journal}{Phys. Rev. B} \textbf{\bibinfo{volume}{55}},
  \bibinfo{pages}{7464} (\bibinfo{year}{1997}),
  \urlprefix\url{https://link.aps.org/doi/10.1103/PhysRevB.55.7464}.

\bibitem[{\citenamefont{An and van Leeuwen}(1991)}]{PhysRevB.44.9410}
\bibinfo{author}{\bibfnamefont{G.}~\bibnamefont{An}} \bibnamefont{and}
  \bibinfo{author}{\bibfnamefont{J.~M.~J.} \bibnamefont{van Leeuwen}},
  \bibinfo{journal}{Phys. Rev. B} \textbf{\bibinfo{volume}{44}},
  \bibinfo{pages}{9410} (\bibinfo{year}{1991}),
  \urlprefix\url{https://link.aps.org/doi/10.1103/PhysRevB.44.9410}.

\bibitem[{\citenamefont{van Bemmel et~al.}(1994)\citenamefont{van Bemmel, ten
  Haaf, van Saarloos, van Leeuwen, and An}}]{PhysRevLett.72.2442}
\bibinfo{author}{\bibfnamefont{H.~J.~M.} \bibnamefont{van Bemmel}},
  \bibinfo{author}{\bibfnamefont{D.~F.~B.} \bibnamefont{ten Haaf}},
  \bibinfo{author}{\bibfnamefont{W.}~\bibnamefont{van Saarloos}},
  \bibinfo{author}{\bibfnamefont{J.~M.~J.} \bibnamefont{van Leeuwen}},
  \bibnamefont{and} \bibinfo{author}{\bibfnamefont{G.}~\bibnamefont{An}},
  \bibinfo{journal}{Phys. Rev. Lett.} \textbf{\bibinfo{volume}{72}},
  \bibinfo{pages}{2442} (\bibinfo{year}{1994}),
  \urlprefix\url{https://link.aps.org/doi/10.1103/PhysRevLett.72.2442}.

\bibitem[{\citenamefont{Capriotti et~al.}(1999)\citenamefont{Capriotti,
  Trumper, and Sorella}}]{PhysRevLett.82.3899}
\bibinfo{author}{\bibfnamefont{L.}~\bibnamefont{Capriotti}},
  \bibinfo{author}{\bibfnamefont{A.~E.} \bibnamefont{Trumper}},
  \bibnamefont{and} \bibinfo{author}{\bibfnamefont{S.}~\bibnamefont{Sorella}},
  \bibinfo{journal}{Phys. Rev. Lett.} \textbf{\bibinfo{volume}{82}},
  \bibinfo{pages}{3899} (\bibinfo{year}{1999}),
  \urlprefix\url{https://link.aps.org/doi/10.1103/PhysRevLett.82.3899}.

\bibitem[{\citenamefont{Chang and Zhang}(2008)}]{PhysRevB.78.165101}
\bibinfo{author}{\bibfnamefont{C.-C.} \bibnamefont{Chang}} \bibnamefont{and}
  \bibinfo{author}{\bibfnamefont{S.}~\bibnamefont{Zhang}},
  \bibinfo{journal}{Phys. Rev. B} \textbf{\bibinfo{volume}{78}},
  \bibinfo{pages}{165101} (\bibinfo{year}{2008}),
  \urlprefix\url{https://link.aps.org/doi/10.1103/PhysRevB.78.165101}.

\bibitem[{\citenamefont{Qin et~al.}(2016)\citenamefont{Qin, Shi, and
  Zhang}}]{PhysRevB.94.235119}
\bibinfo{author}{\bibfnamefont{M.}~\bibnamefont{Qin}},
  \bibinfo{author}{\bibfnamefont{H.}~\bibnamefont{Shi}}, \bibnamefont{and}
  \bibinfo{author}{\bibfnamefont{S.}~\bibnamefont{Zhang}},
  \bibinfo{journal}{Phys. Rev. B} \textbf{\bibinfo{volume}{94}},
  \bibinfo{pages}{235119} (\bibinfo{year}{2016}),
  \urlprefix\url{https://link.aps.org/doi/10.1103/PhysRevB.94.235119}.

\bibitem[{\citenamefont{Chandra and Doucot}(1988)}]{PhysRevB.38.9335}
\bibinfo{author}{\bibfnamefont{P.}~\bibnamefont{Chandra}} \bibnamefont{and}
  \bibinfo{author}{\bibfnamefont{B.}~\bibnamefont{Doucot}},
  \bibinfo{journal}{Phys. Rev. B} \textbf{\bibinfo{volume}{38}},
  \bibinfo{pages}{9335} (\bibinfo{year}{1988}),
  \urlprefix\url{https://link.aps.org/doi/10.1103/PhysRevB.38.9335}.

\bibitem[{\citenamefont{Dagotto and Moreo}(1989)}]{PhysRevLett.63.2148}
\bibinfo{author}{\bibfnamefont{E.}~\bibnamefont{Dagotto}} \bibnamefont{and}
  \bibinfo{author}{\bibfnamefont{A.}~\bibnamefont{Moreo}},
  \bibinfo{journal}{Phys. Rev. Lett.} \textbf{\bibinfo{volume}{63}},
  \bibinfo{pages}{2148} (\bibinfo{year}{1989}),
  \urlprefix\url{https://link.aps.org/doi/10.1103/PhysRevLett.63.2148}.

\bibitem[{\citenamefont{Capriotti and Sorella}(2000)}]{PhysRevLett.84.3173}
\bibinfo{author}{\bibfnamefont{L.}~\bibnamefont{Capriotti}} \bibnamefont{and}
  \bibinfo{author}{\bibfnamefont{S.}~\bibnamefont{Sorella}},
  \bibinfo{journal}{Phys. Rev. Lett.} \textbf{\bibinfo{volume}{84}},
  \bibinfo{pages}{3173} (\bibinfo{year}{2000}),
  \urlprefix\url{https://link.aps.org/doi/10.1103/PhysRevLett.84.3173}.

\bibitem[{\citenamefont{Sandvik}(1997)}]{PhysRevB.56.11678}
\bibinfo{author}{\bibfnamefont{A.~W.} \bibnamefont{Sandvik}},
  \bibinfo{journal}{Phys. Rev. B} \textbf{\bibinfo{volume}{56}},
  \bibinfo{pages}{11678} (\bibinfo{year}{1997}),
  \urlprefix\url{https://link.aps.org/doi/10.1103/PhysRevB.56.11678}.

\bibitem[{\citenamefont{Jiang and Wiese}(2011)}]{PhysRevB.83.155120}
\bibinfo{author}{\bibfnamefont{F.-J.} \bibnamefont{Jiang}} \bibnamefont{and}
  \bibinfo{author}{\bibfnamefont{U.-J.} \bibnamefont{Wiese}},
  \bibinfo{journal}{Phys. Rev. B} \textbf{\bibinfo{volume}{83}},
  \bibinfo{pages}{155120} (\bibinfo{year}{2011}),
  \urlprefix\url{https://link.aps.org/doi/10.1103/PhysRevB.83.155120}.

\bibitem[{\citenamefont{Sandvik and Evertz}(2010)}]{PhysRevB.82.024407}
\bibinfo{author}{\bibfnamefont{A.~W.} \bibnamefont{Sandvik}} \bibnamefont{and}
  \bibinfo{author}{\bibfnamefont{H.~G.} \bibnamefont{Evertz}},
  \bibinfo{journal}{Phys. Rev. B} \textbf{\bibinfo{volume}{82}},
  \bibinfo{pages}{024407} (\bibinfo{year}{2010}),
  \urlprefix\url{https://link.aps.org/doi/10.1103/PhysRevB.82.024407}.

\bibitem[{\citenamefont{Mambrini et~al.}(2006)\citenamefont{Mambrini,
  L\"auchli, Poilblanc, and Mila}}]{PhysRevB.74.144422}
\bibinfo{author}{\bibfnamefont{M.}~\bibnamefont{Mambrini}},
  \bibinfo{author}{\bibfnamefont{A.}~\bibnamefont{L\"auchli}},
  \bibinfo{author}{\bibfnamefont{D.}~\bibnamefont{Poilblanc}},
  \bibnamefont{and} \bibinfo{author}{\bibfnamefont{F.}~\bibnamefont{Mila}},
  \bibinfo{journal}{Phys. Rev. B} \textbf{\bibinfo{volume}{74}},
  \bibinfo{pages}{144422} (\bibinfo{year}{2006}),
  \urlprefix\url{https://link.aps.org/doi/10.1103/PhysRevB.74.144422}.

\bibitem[{\citenamefont{Murg et~al.}(2009)\citenamefont{Murg, Verstraete, and
  Cirac}}]{PhysRevB.79.195119}
\bibinfo{author}{\bibfnamefont{V.}~\bibnamefont{Murg}},
  \bibinfo{author}{\bibfnamefont{F.}~\bibnamefont{Verstraete}},
  \bibnamefont{and} \bibinfo{author}{\bibfnamefont{J.~I.} \bibnamefont{Cirac}},
  \bibinfo{journal}{Phys. Rev. B} \textbf{\bibinfo{volume}{79}},
  \bibinfo{pages}{195119} (\bibinfo{year}{2009}),
  \urlprefix\url{https://link.aps.org/doi/10.1103/PhysRevB.79.195119}.

\bibitem[{\citenamefont{Yu and Kao}(2012)}]{PhysRevB.85.094407}
\bibinfo{author}{\bibfnamefont{J.-F.} \bibnamefont{Yu}} \bibnamefont{and}
  \bibinfo{author}{\bibfnamefont{Y.-J.} \bibnamefont{Kao}},
  \bibinfo{journal}{Phys. Rev. B} \textbf{\bibinfo{volume}{85}},
  \bibinfo{pages}{094407} (\bibinfo{year}{2012}),
  \urlprefix\url{https://link.aps.org/doi/10.1103/PhysRevB.85.094407}.

\bibitem[{\citenamefont{Jiang et~al.}(2012)\citenamefont{Jiang, Yao, and
  Balents}}]{PhysRevB.86.024424}
\bibinfo{author}{\bibfnamefont{H.-C.} \bibnamefont{Jiang}},
  \bibinfo{author}{\bibfnamefont{H.}~\bibnamefont{Yao}}, \bibnamefont{and}
  \bibinfo{author}{\bibfnamefont{L.}~\bibnamefont{Balents}},
  \bibinfo{journal}{Phys. Rev. B} \textbf{\bibinfo{volume}{86}},
  \bibinfo{pages}{024424} (\bibinfo{year}{2012}),
  \urlprefix\url{https://link.aps.org/doi/10.1103/PhysRevB.86.024424}.

\bibitem[{\citenamefont{Mezzacapo}(2012)}]{PhysRevB.86.045115}
\bibinfo{author}{\bibfnamefont{F.}~\bibnamefont{Mezzacapo}},
  \bibinfo{journal}{Phys. Rev. B} \textbf{\bibinfo{volume}{86}},
  \bibinfo{pages}{045115} (\bibinfo{year}{2012}),
  \urlprefix\url{https://link.aps.org/doi/10.1103/PhysRevB.86.045115}.

\bibitem[{\citenamefont{Hu et~al.}(2013)\citenamefont{Hu, Becca, Parola, and
  Sorella}}]{PhysRevB.88.060402}
\bibinfo{author}{\bibfnamefont{W.-J.} \bibnamefont{Hu}},
  \bibinfo{author}{\bibfnamefont{F.}~\bibnamefont{Becca}},
  \bibinfo{author}{\bibfnamefont{A.}~\bibnamefont{Parola}}, \bibnamefont{and}
  \bibinfo{author}{\bibfnamefont{S.}~\bibnamefont{Sorella}},
  \bibinfo{journal}{Phys. Rev. B} \textbf{\bibinfo{volume}{88}},
  \bibinfo{pages}{060402} (\bibinfo{year}{2013}),
  \urlprefix\url{https://link.aps.org/doi/10.1103/PhysRevB.88.060402}.

\bibitem[{\citenamefont{Gong et~al.}(2014)\citenamefont{Gong, Zhu, Sheng,
  Motrunich, and Fisher}}]{PhysRevLett.113.027201}
\bibinfo{author}{\bibfnamefont{S.-S.} \bibnamefont{Gong}},
  \bibinfo{author}{\bibfnamefont{W.}~\bibnamefont{Zhu}},
  \bibinfo{author}{\bibfnamefont{D.~N.} \bibnamefont{Sheng}},
  \bibinfo{author}{\bibfnamefont{O.~I.} \bibnamefont{Motrunich}},
  \bibnamefont{and} \bibinfo{author}{\bibfnamefont{M.~P.~A.}
  \bibnamefont{Fisher}}, \bibinfo{journal}{Phys. Rev. Lett.}
  \textbf{\bibinfo{volume}{113}}, \bibinfo{pages}{027201}
  (\bibinfo{year}{2014}),
  \urlprefix\url{https://link.aps.org/doi/10.1103/PhysRevLett.113.027201}.

\bibitem[{\citenamefont{Morita et~al.}(2015)\citenamefont{Morita, Kaneko, and
  Imada}}]{doi:10.7566/JPSJ.84.024720}
\bibinfo{author}{\bibfnamefont{S.}~\bibnamefont{Morita}},
  \bibinfo{author}{\bibfnamefont{R.}~\bibnamefont{Kaneko}}, \bibnamefont{and}
  \bibinfo{author}{\bibfnamefont{M.}~\bibnamefont{Imada}},
  \bibinfo{journal}{Journal of the Physical Society of Japan}
  \textbf{\bibinfo{volume}{84}}, \bibinfo{pages}{024720}
  (\bibinfo{year}{2015}), \eprint{https://doi.org/10.7566/JPSJ.84.024720},
  \urlprefix\url{https://doi.org/10.7566/JPSJ.84.024720}.

\bibitem[{\citenamefont{Wang and Sandvik}(2018)}]{PhysRevLett.121.107202}
\bibinfo{author}{\bibfnamefont{L.}~\bibnamefont{Wang}} \bibnamefont{and}
  \bibinfo{author}{\bibfnamefont{A.~W.} \bibnamefont{Sandvik}},
  \bibinfo{journal}{Phys. Rev. Lett.} \textbf{\bibinfo{volume}{121}},
  \bibinfo{pages}{107202} (\bibinfo{year}{2018}),
  \urlprefix\url{https://link.aps.org/doi/10.1103/PhysRevLett.121.107202}.

\bibitem[{\citenamefont{Suzuki}(1976)}]{10.1143/PTP.56.1454}
\bibinfo{author}{\bibfnamefont{M.}~\bibnamefont{Suzuki}},
  \bibinfo{journal}{Progress of Theoretical Physics}
  \textbf{\bibinfo{volume}{56}}, \bibinfo{pages}{1454} (\bibinfo{year}{1976}),
  ISSN \bibinfo{issn}{0033-068X},
  \eprint{https://academic.oup.com/ptp/article-pdf/56/5/1454/5264429/56-5-1454.pdf},
  \urlprefix\url{https://doi.org/10.1143/PTP.56.1454}.

\bibitem[{\citenamefont{Pirvu et~al.}(2010)\citenamefont{Pirvu, Murg, Cirac,
  and Verstraete}}]{Pirvu_2010}
\bibinfo{author}{\bibfnamefont{B.}~\bibnamefont{Pirvu}},
  \bibinfo{author}{\bibfnamefont{V.}~\bibnamefont{Murg}},
  \bibinfo{author}{\bibfnamefont{J.~I.} \bibnamefont{Cirac}}, \bibnamefont{and}
  \bibinfo{author}{\bibfnamefont{F.}~\bibnamefont{Verstraete}},
  \bibinfo{journal}{New Journal of Physics} \textbf{\bibinfo{volume}{12}},
  \bibinfo{pages}{025012} (\bibinfo{year}{2010}),
  \urlprefix\url{https://doi.org/10.1088%2F1367-2630%2F12%2F2%2F025012}.

\bibitem[{\citenamefont{Evenbly}(2018)}]{PhysRevB.98.085155}
\bibinfo{author}{\bibfnamefont{G.}~\bibnamefont{Evenbly}},
  \bibinfo{journal}{Phys. Rev. B} \textbf{\bibinfo{volume}{98}},
  \bibinfo{pages}{085155} (\bibinfo{year}{2018}),
  \urlprefix\url{https://link.aps.org/doi/10.1103/PhysRevB.98.085155}.

\bibitem[{\citenamefont{{Wang} and {Verstraete}}(2011)}]{2011arXiv1110.4362W}
\bibinfo{author}{\bibfnamefont{L.}~\bibnamefont{{Wang}}} \bibnamefont{and}
  \bibinfo{author}{\bibfnamefont{F.}~\bibnamefont{{Verstraete}}},
  \bibinfo{journal}{arXiv e-prints} \bibinfo{eid}{arXiv:1110.4362}
  (\bibinfo{year}{2011}), \eprint{1110.4362}.

\bibitem[{\citenamefont{Xie et~al.}(2017)\citenamefont{Xie, Liao, Huang, Xie,
  Chen, Liu, and Xiang}}]{PhysRevB.96.045128}
\bibinfo{author}{\bibfnamefont{Z.~Y.} \bibnamefont{Xie}},
  \bibinfo{author}{\bibfnamefont{H.~J.} \bibnamefont{Liao}},
  \bibinfo{author}{\bibfnamefont{R.~Z.} \bibnamefont{Huang}},
  \bibinfo{author}{\bibfnamefont{H.~D.} \bibnamefont{Xie}},
  \bibinfo{author}{\bibfnamefont{J.}~\bibnamefont{Chen}},
  \bibinfo{author}{\bibfnamefont{Z.~Y.} \bibnamefont{Liu}}, \bibnamefont{and}
  \bibinfo{author}{\bibfnamefont{T.}~\bibnamefont{Xiang}},
  \bibinfo{journal}{Phys. Rev. B} \textbf{\bibinfo{volume}{96}},
  \bibinfo{pages}{045128} (\bibinfo{year}{2017}),
  \urlprefix\url{https://link.aps.org/doi/10.1103/PhysRevB.96.045128}.

\bibitem[{\citenamefont{Corboz et~al.}(2010)\citenamefont{Corboz, Or\'us,
  Bauer, and Vidal}}]{PhysRevB.81.165104}
\bibinfo{author}{\bibfnamefont{P.}~\bibnamefont{Corboz}},
  \bibinfo{author}{\bibfnamefont{R.}~\bibnamefont{Or\'us}},
  \bibinfo{author}{\bibfnamefont{B.}~\bibnamefont{Bauer}}, \bibnamefont{and}
  \bibinfo{author}{\bibfnamefont{G.}~\bibnamefont{Vidal}},
  \bibinfo{journal}{Phys. Rev. B} \textbf{\bibinfo{volume}{81}},
  \bibinfo{pages}{165104} (\bibinfo{year}{2010}),
  \urlprefix\url{https://link.aps.org/doi/10.1103/PhysRevB.81.165104}.

\bibitem[{\citenamefont{Carlson}(1989)}]{PhysRevB.40.846}
\bibinfo{author}{\bibfnamefont{J.}~\bibnamefont{Carlson}},
  \bibinfo{journal}{Phys. Rev. B} \textbf{\bibinfo{volume}{40}},
  \bibinfo{pages}{846} (\bibinfo{year}{1989}),
  \urlprefix\url{https://link.aps.org/doi/10.1103/PhysRevB.40.846}.

\bibitem[{\citenamefont{Trivedi and Ceperley}(1989)}]{PhysRevB.40.2737}
\bibinfo{author}{\bibfnamefont{N.}~\bibnamefont{Trivedi}} \bibnamefont{and}
  \bibinfo{author}{\bibfnamefont{D.~M.} \bibnamefont{Ceperley}},
  \bibinfo{journal}{Phys. Rev. B} \textbf{\bibinfo{volume}{40}},
  \bibinfo{pages}{2737} (\bibinfo{year}{1989}),
  \urlprefix\url{https://link.aps.org/doi/10.1103/PhysRevB.40.2737}.

\bibitem[{\citenamefont{Trivedi and Ceperley}(1990)}]{PhysRevB.41.4552}
\bibinfo{author}{\bibfnamefont{N.}~\bibnamefont{Trivedi}} \bibnamefont{and}
  \bibinfo{author}{\bibfnamefont{D.~M.} \bibnamefont{Ceperley}},
  \bibinfo{journal}{Phys. Rev. B} \textbf{\bibinfo{volume}{41}},
  \bibinfo{pages}{4552} (\bibinfo{year}{1990}),
  \urlprefix\url{https://link.aps.org/doi/10.1103/PhysRevB.41.4552}.

\bibitem[{\citenamefont{Runge}(1992)}]{PhysRevB.45.7229}
\bibinfo{author}{\bibfnamefont{K.~J.} \bibnamefont{Runge}},
  \bibinfo{journal}{Phys. Rev. B} \textbf{\bibinfo{volume}{45}},
  \bibinfo{pages}{7229} (\bibinfo{year}{1992}),
  \urlprefix\url{https://link.aps.org/doi/10.1103/PhysRevB.45.7229}.

\bibitem[{\citenamefont{ten Haaf et~al.}(1995)\citenamefont{ten Haaf, van
  Bemmel, van Leeuwen, van Saarloos, and Ceperley}}]{PhysRevB.51.13039}
\bibinfo{author}{\bibfnamefont{D.~F.~B.} \bibnamefont{ten Haaf}},
  \bibinfo{author}{\bibfnamefont{H.~J.~M.} \bibnamefont{van Bemmel}},
  \bibinfo{author}{\bibfnamefont{J.~M.~J.} \bibnamefont{van Leeuwen}},
  \bibinfo{author}{\bibfnamefont{W.}~\bibnamefont{van Saarloos}},
  \bibnamefont{and} \bibinfo{author}{\bibfnamefont{D.~M.}
  \bibnamefont{Ceperley}}, \bibinfo{journal}{Phys. Rev. B}
  \textbf{\bibinfo{volume}{51}}, \bibinfo{pages}{13039} (\bibinfo{year}{1995}),
  \urlprefix\url{https://link.aps.org/doi/10.1103/PhysRevB.51.13039}.

\bibitem[{\citenamefont{Calandra~Buonaura and
  Sorella}(1998)}]{PhysRevB.57.11446}
\bibinfo{author}{\bibfnamefont{M.}~\bibnamefont{Calandra~Buonaura}}
  \bibnamefont{and} \bibinfo{author}{\bibfnamefont{S.}~\bibnamefont{Sorella}},
  \bibinfo{journal}{Phys. Rev. B} \textbf{\bibinfo{volume}{57}},
  \bibinfo{pages}{11446} (\bibinfo{year}{1998}),
  \urlprefix\url{https://link.aps.org/doi/10.1103/PhysRevB.57.11446}.

\bibitem[{\citenamefont{Schuch et~al.}(2007)\citenamefont{Schuch, Wolf,
  Verstraete, and Cirac}}]{PhysRevLett.98.140506}
\bibinfo{author}{\bibfnamefont{N.}~\bibnamefont{Schuch}},
  \bibinfo{author}{\bibfnamefont{M.~M.} \bibnamefont{Wolf}},
  \bibinfo{author}{\bibfnamefont{F.}~\bibnamefont{Verstraete}},
  \bibnamefont{and} \bibinfo{author}{\bibfnamefont{J.~I.} \bibnamefont{Cirac}},
  \bibinfo{journal}{Phys. Rev. Lett.} \textbf{\bibinfo{volume}{98}},
  \bibinfo{pages}{140506} (\bibinfo{year}{2007}),
  \urlprefix\url{https://link.aps.org/doi/10.1103/PhysRevLett.98.140506}.

\bibitem[{\citenamefont{Nishino and Okunishi}(1996)}]{doi:10.1143/JPSJ.65.891}
\bibinfo{author}{\bibfnamefont{T.}~\bibnamefont{Nishino}} \bibnamefont{and}
  \bibinfo{author}{\bibfnamefont{K.}~\bibnamefont{Okunishi}},
  \bibinfo{journal}{Journal of the Physical Society of Japan}
  \textbf{\bibinfo{volume}{65}}, \bibinfo{pages}{891} (\bibinfo{year}{1996}),
  \eprint{https://doi.org/10.1143/JPSJ.65.891},
  \urlprefix\url{https://doi.org/10.1143/JPSJ.65.891}.

\bibitem[{\citenamefont{Levin and Nave}(2007)}]{PhysRevLett.99.120601}
\bibinfo{author}{\bibfnamefont{M.}~\bibnamefont{Levin}} \bibnamefont{and}
  \bibinfo{author}{\bibfnamefont{C.~P.} \bibnamefont{Nave}},
  \bibinfo{journal}{Phys. Rev. Lett.} \textbf{\bibinfo{volume}{99}},
  \bibinfo{pages}{120601} (\bibinfo{year}{2007}),
  \urlprefix\url{https://link.aps.org/doi/10.1103/PhysRevLett.99.120601}.

\bibitem[{\citenamefont{Gu and Wen}(2009)}]{PhysRevB.80.155131}
\bibinfo{author}{\bibfnamefont{Z.-C.} \bibnamefont{Gu}} \bibnamefont{and}
  \bibinfo{author}{\bibfnamefont{X.-G.} \bibnamefont{Wen}},
  \bibinfo{journal}{Phys. Rev. B} \textbf{\bibinfo{volume}{80}},
  \bibinfo{pages}{155131} (\bibinfo{year}{2009}),
  \urlprefix\url{https://link.aps.org/doi/10.1103/PhysRevB.80.155131}.

\bibitem[{\citenamefont{Zhao et~al.}(2010)\citenamefont{Zhao, Xie, Chen, Wei,
  Cai, and Xiang}}]{PhysRevB.81.174411}
\bibinfo{author}{\bibfnamefont{H.~H.} \bibnamefont{Zhao}},
  \bibinfo{author}{\bibfnamefont{Z.~Y.} \bibnamefont{Xie}},
  \bibinfo{author}{\bibfnamefont{Q.~N.} \bibnamefont{Chen}},
  \bibinfo{author}{\bibfnamefont{Z.~C.} \bibnamefont{Wei}},
  \bibinfo{author}{\bibfnamefont{J.~W.} \bibnamefont{Cai}}, \bibnamefont{and}
  \bibinfo{author}{\bibfnamefont{T.}~\bibnamefont{Xiang}},
  \bibinfo{journal}{Phys. Rev. B} \textbf{\bibinfo{volume}{81}},
  \bibinfo{pages}{174411} (\bibinfo{year}{2010}),
  \urlprefix\url{https://link.aps.org/doi/10.1103/PhysRevB.81.174411}.

\bibitem[{\citenamefont{Lubasch et~al.}(2014)\citenamefont{Lubasch, Cirac, and
  Ba\~nuls}}]{PhysRevB.90.064425}
\bibinfo{author}{\bibfnamefont{M.}~\bibnamefont{Lubasch}},
  \bibinfo{author}{\bibfnamefont{J.~I.} \bibnamefont{Cirac}}, \bibnamefont{and}
  \bibinfo{author}{\bibfnamefont{M.-C.} \bibnamefont{Ba\~nuls}},
  \bibinfo{journal}{Phys. Rev. B} \textbf{\bibinfo{volume}{90}},
  \bibinfo{pages}{064425} (\bibinfo{year}{2014}),
  \urlprefix\url{https://link.aps.org/doi/10.1103/PhysRevB.90.064425}.

\bibitem[{\citenamefont{Xie et~al.}(2012)\citenamefont{Xie, Chen, Qin, Zhu,
  Yang, and Xiang}}]{PhysRevB.86.045139}
\bibinfo{author}{\bibfnamefont{Z.~Y.} \bibnamefont{Xie}},
  \bibinfo{author}{\bibfnamefont{J.}~\bibnamefont{Chen}},
  \bibinfo{author}{\bibfnamefont{M.~P.} \bibnamefont{Qin}},
  \bibinfo{author}{\bibfnamefont{J.~W.} \bibnamefont{Zhu}},
  \bibinfo{author}{\bibfnamefont{L.~P.} \bibnamefont{Yang}}, \bibnamefont{and}
  \bibinfo{author}{\bibfnamefont{T.}~\bibnamefont{Xiang}},
  \bibinfo{journal}{Phys. Rev. B} \textbf{\bibinfo{volume}{86}},
  \bibinfo{pages}{045139} (\bibinfo{year}{2012}),
  \urlprefix\url{https://link.aps.org/doi/10.1103/PhysRevB.86.045139}.

\bibitem[{\citenamefont{Evenbly and Vidal}(2015)}]{PhysRevLett.115.180405}
\bibinfo{author}{\bibfnamefont{G.}~\bibnamefont{Evenbly}} \bibnamefont{and}
  \bibinfo{author}{\bibfnamefont{G.}~\bibnamefont{Vidal}},
  \bibinfo{journal}{Phys. Rev. Lett.} \textbf{\bibinfo{volume}{115}},
  \bibinfo{pages}{180405} (\bibinfo{year}{2015}),
  \urlprefix\url{https://link.aps.org/doi/10.1103/PhysRevLett.115.180405}.

\bibitem[{\citenamefont{Yang et~al.}(2017)\citenamefont{Yang, Gu, and
  Wen}}]{PhysRevLett.118.110504}
\bibinfo{author}{\bibfnamefont{S.}~\bibnamefont{Yang}},
  \bibinfo{author}{\bibfnamefont{Z.-C.} \bibnamefont{Gu}}, \bibnamefont{and}
  \bibinfo{author}{\bibfnamefont{X.-G.} \bibnamefont{Wen}},
  \bibinfo{journal}{Phys. Rev. Lett.} \textbf{\bibinfo{volume}{118}},
  \bibinfo{pages}{110504} (\bibinfo{year}{2017}),
  \urlprefix\url{https://link.aps.org/doi/10.1103/PhysRevLett.118.110504}.

\bibitem[{\citenamefont{Sandvik and Vidal}(2007)}]{PhysRevLett.99.220602}
\bibinfo{author}{\bibfnamefont{A.~W.} \bibnamefont{Sandvik}} \bibnamefont{and}
  \bibinfo{author}{\bibfnamefont{G.}~\bibnamefont{Vidal}},
  \bibinfo{journal}{Phys. Rev. Lett.} \textbf{\bibinfo{volume}{99}},
  \bibinfo{pages}{220602} (\bibinfo{year}{2007}),
  \urlprefix\url{https://link.aps.org/doi/10.1103/PhysRevLett.99.220602}.

\bibitem[{\citenamefont{Schuch et~al.}(2008)\citenamefont{Schuch, Wolf,
  Verstraete, and Cirac}}]{PhysRevLett.100.040501}
\bibinfo{author}{\bibfnamefont{N.}~\bibnamefont{Schuch}},
  \bibinfo{author}{\bibfnamefont{M.~M.} \bibnamefont{Wolf}},
  \bibinfo{author}{\bibfnamefont{F.}~\bibnamefont{Verstraete}},
  \bibnamefont{and} \bibinfo{author}{\bibfnamefont{J.~I.} \bibnamefont{Cirac}},
  \bibinfo{journal}{Phys. Rev. Lett.} \textbf{\bibinfo{volume}{100}},
  \bibinfo{pages}{040501} (\bibinfo{year}{2008}),
  \urlprefix\url{https://link.aps.org/doi/10.1103/PhysRevLett.100.040501}.

\bibitem[{\citenamefont{Wang et~al.}(2011)\citenamefont{Wang,
  Pi\ifmmode~\check{z}\else \v{z}\fi{}orn, and
  Verstraete}}]{PhysRevB.83.134421}
\bibinfo{author}{\bibfnamefont{L.}~\bibnamefont{Wang}},
  \bibinfo{author}{\bibfnamefont{I.}~\bibnamefont{Pi\ifmmode~\check{z}\else
  \v{z}\fi{}orn}}, \bibnamefont{and}
  \bibinfo{author}{\bibfnamefont{F.}~\bibnamefont{Verstraete}},
  \bibinfo{journal}{Phys. Rev. B} \textbf{\bibinfo{volume}{83}},
  \bibinfo{pages}{134421} (\bibinfo{year}{2011}),
  \urlprefix\url{https://link.aps.org/doi/10.1103/PhysRevB.83.134421}.

\bibitem[{\citenamefont{Sikora et~al.}(2015)\citenamefont{Sikora, Chang, Chou,
  Pollmann, and Kao}}]{PhysRevB.91.165113}
\bibinfo{author}{\bibfnamefont{O.}~\bibnamefont{Sikora}},
  \bibinfo{author}{\bibfnamefont{H.-W.} \bibnamefont{Chang}},
  \bibinfo{author}{\bibfnamefont{C.-P.} \bibnamefont{Chou}},
  \bibinfo{author}{\bibfnamefont{F.}~\bibnamefont{Pollmann}}, \bibnamefont{and}
  \bibinfo{author}{\bibfnamefont{Y.-J.} \bibnamefont{Kao}},
  \bibinfo{journal}{Phys. Rev. B} \textbf{\bibinfo{volume}{91}},
  \bibinfo{pages}{165113} (\bibinfo{year}{2015}),
  \urlprefix\url{https://link.aps.org/doi/10.1103/PhysRevB.91.165113}.

\bibitem[{\citenamefont{Liu et~al.}(2017)\citenamefont{Liu, Dong, Han, Guo, and
  He}}]{PhysRevB.95.195154}
\bibinfo{author}{\bibfnamefont{W.-Y.} \bibnamefont{Liu}},
  \bibinfo{author}{\bibfnamefont{S.-J.} \bibnamefont{Dong}},
  \bibinfo{author}{\bibfnamefont{Y.-J.} \bibnamefont{Han}},
  \bibinfo{author}{\bibfnamefont{G.-C.} \bibnamefont{Guo}}, \bibnamefont{and}
  \bibinfo{author}{\bibfnamefont{L.}~\bibnamefont{He}}, \bibinfo{journal}{Phys.
  Rev. B} \textbf{\bibinfo{volume}{95}}, \bibinfo{pages}{195154}
  (\bibinfo{year}{2017}),
  \urlprefix\url{https://link.aps.org/doi/10.1103/PhysRevB.95.195154}.

\bibitem[{\citenamefont{Zhao et~al.}(2017)\citenamefont{Zhao, Ido, Morita, and
  Imada}}]{PhysRevB.96.085103}
\bibinfo{author}{\bibfnamefont{H.-H.} \bibnamefont{Zhao}},
  \bibinfo{author}{\bibfnamefont{K.}~\bibnamefont{Ido}},
  \bibinfo{author}{\bibfnamefont{S.}~\bibnamefont{Morita}}, \bibnamefont{and}
  \bibinfo{author}{\bibfnamefont{M.}~\bibnamefont{Imada}},
  \bibinfo{journal}{Phys. Rev. B} \textbf{\bibinfo{volume}{96}},
  \bibinfo{pages}{085103} (\bibinfo{year}{2017}),
  \urlprefix\url{https://link.aps.org/doi/10.1103/PhysRevB.96.085103}.

\bibitem[{\citenamefont{du~Croo~de Jongh et~al.}(2000)\citenamefont{du~Croo~de
  Jongh, van Leeuwen, and van Saarloos}}]{PhysRevB.62.14844}
\bibinfo{author}{\bibfnamefont{M.~S.~L.} \bibnamefont{du~Croo~de Jongh}},
  \bibinfo{author}{\bibfnamefont{J.~M.~J.} \bibnamefont{van Leeuwen}},
  \bibnamefont{and} \bibinfo{author}{\bibfnamefont{W.}~\bibnamefont{van
  Saarloos}}, \bibinfo{journal}{Phys. Rev. B} \textbf{\bibinfo{volume}{62}},
  \bibinfo{pages}{14844} (\bibinfo{year}{2000}),
  \urlprefix\url{https://link.aps.org/doi/10.1103/PhysRevB.62.14844}.

\bibitem[{\citenamefont{Wouters et~al.}(2014)\citenamefont{Wouters, Verstichel,
  Van~Neck, and Chan}}]{PhysRevB.90.045104}
\bibinfo{author}{\bibfnamefont{S.}~\bibnamefont{Wouters}},
  \bibinfo{author}{\bibfnamefont{B.}~\bibnamefont{Verstichel}},
  \bibinfo{author}{\bibfnamefont{D.}~\bibnamefont{Van~Neck}}, \bibnamefont{and}
  \bibinfo{author}{\bibfnamefont{G.~K.-L.} \bibnamefont{Chan}},
  \bibinfo{journal}{Phys. Rev. B} \textbf{\bibinfo{volume}{90}},
  \bibinfo{pages}{045104} (\bibinfo{year}{2014}),
  \urlprefix\url{https://link.aps.org/doi/10.1103/PhysRevB.90.045104}.

\bibitem[{\citenamefont{Marshall and
  Peierls}(1955)}]{doi:10.1098/rspa.1955.0200}
\bibinfo{author}{\bibfnamefont{W.}~\bibnamefont{Marshall}} \bibnamefont{and}
  \bibinfo{author}{\bibfnamefont{R.~E.} \bibnamefont{Peierls}},
  \bibinfo{journal}{Proceedings of the Royal Society of London. Series A.
  Mathematical and Physical Sciences} \textbf{\bibinfo{volume}{232}},
  \bibinfo{pages}{48} (\bibinfo{year}{1955}),
  \eprint{https://royalsocietypublishing.org/doi/pdf/10.1098/rspa.1955.0200},
  \urlprefix\url{https://royalsocietypublishing.org/doi/abs/10.1098/rspa.1955.0200}.

\bibitem[{\citenamefont{Sorella and Capriotti}(2000)}]{PhysRevB.61.2599}
\bibinfo{author}{\bibfnamefont{S.}~\bibnamefont{Sorella}} \bibnamefont{and}
  \bibinfo{author}{\bibfnamefont{L.}~\bibnamefont{Capriotti}},
  \bibinfo{journal}{Phys. Rev. B} \textbf{\bibinfo{volume}{61}},
  \bibinfo{pages}{2599} (\bibinfo{year}{2000}),
  \urlprefix\url{https://link.aps.org/doi/10.1103/PhysRevB.61.2599}.

\bibitem[{\citenamefont{Huang et~al.}(2018)\citenamefont{Huang, Liao, Liu, Xie,
  Xie, Zhao, Chen, and Xiang}}]{Huang_2018}
\bibinfo{author}{\bibfnamefont{R.-Z.} \bibnamefont{Huang}},
  \bibinfo{author}{\bibfnamefont{H.-J.} \bibnamefont{Liao}},
  \bibinfo{author}{\bibfnamefont{Z.-Y.} \bibnamefont{Liu}},
  \bibinfo{author}{\bibfnamefont{H.-D.} \bibnamefont{Xie}},
  \bibinfo{author}{\bibfnamefont{Z.-Y.} \bibnamefont{Xie}},
  \bibinfo{author}{\bibfnamefont{H.-H.} \bibnamefont{Zhao}},
  \bibinfo{author}{\bibfnamefont{J.}~\bibnamefont{Chen}}, \bibnamefont{and}
  \bibinfo{author}{\bibfnamefont{T.}~\bibnamefont{Xiang}},
  \bibinfo{journal}{Chinese Physics B} \textbf{\bibinfo{volume}{27}},
  \bibinfo{pages}{070501} (\bibinfo{year}{2018}),
  \urlprefix\url{https://doi.org/10.1088%2F1674-1056%2F27%2F7%2F070501}.

\bibitem[{\citenamefont{Shi et~al.}(2006)\citenamefont{Shi, Duan, and
  Vidal}}]{PhysRevA.74.022320}
\bibinfo{author}{\bibfnamefont{Y.-Y.} \bibnamefont{Shi}},
  \bibinfo{author}{\bibfnamefont{L.-M.} \bibnamefont{Duan}}, \bibnamefont{and}
  \bibinfo{author}{\bibfnamefont{G.}~\bibnamefont{Vidal}},
  \bibinfo{journal}{Phys. Rev. A} \textbf{\bibinfo{volume}{74}},
  \bibinfo{pages}{022320} (\bibinfo{year}{2006}),
  \urlprefix\url{https://link.aps.org/doi/10.1103/PhysRevA.74.022320}.

\bibitem[{\citenamefont{Vidal}(2008)}]{PhysRevLett.101.110501}
\bibinfo{author}{\bibfnamefont{G.}~\bibnamefont{Vidal}},
  \bibinfo{journal}{Phys. Rev. Lett.} \textbf{\bibinfo{volume}{101}},
  \bibinfo{pages}{110501} (\bibinfo{year}{2008}),
  \urlprefix\url{https://link.aps.org/doi/10.1103/PhysRevLett.101.110501}.

\bibitem[{\citenamefont{Xie et~al.}(2014)\citenamefont{Xie, Chen, Yu, Kong,
  Normand, and Xiang}}]{PhysRevX.4.011025}
\bibinfo{author}{\bibfnamefont{Z.~Y.} \bibnamefont{Xie}},
  \bibinfo{author}{\bibfnamefont{J.}~\bibnamefont{Chen}},
  \bibinfo{author}{\bibfnamefont{J.~F.} \bibnamefont{Yu}},
  \bibinfo{author}{\bibfnamefont{X.}~\bibnamefont{Kong}},
  \bibinfo{author}{\bibfnamefont{B.}~\bibnamefont{Normand}}, \bibnamefont{and}
  \bibinfo{author}{\bibfnamefont{T.}~\bibnamefont{Xiang}},
  \bibinfo{journal}{Phys. Rev. X} \textbf{\bibinfo{volume}{4}},
  \bibinfo{pages}{011025} (\bibinfo{year}{2014}),
  \urlprefix\url{https://link.aps.org/doi/10.1103/PhysRevX.4.011025}.

\end{thebibliography}

\appendix

\end{document}